\documentclass[aps,twocolumn,preprintnumbers,
superscriptaddress,floatfix,nofootinbib]{revtex4}

\usepackage{graphicx,color,dcolumn,booktabs,bm}
\usepackage{amsmath}
\usepackage[caption=false]{subfig}
\usepackage{float}
\usepackage{longtable,lscape}
\usepackage{soul}

\usepackage{txfonts}
\usepackage{mathrsfs}

\usepackage{overpic}

\usepackage{epstopdf}

\usepackage{appendix}
\usepackage{indentfirst}

\usepackage{slashed}

\usepackage{multirow}
\usepackage{array}

\def\mev{\mathrm{MeV}}
\def\fm{\mathrm{fm}}
\def\bseq{\begin{subequations}}
\def\eseq{\end{subequations}}

\usepackage{cellspace}
\graphicspath{{Figures/}}

\usepackage[colorlinks, citecolor=blue,anchorcolor=red,menucolor=red, linkcolor=red,filecolor=red,runcolor=red,urlcolor=blue,frenchlinks=true]{hyperref}
\begin{document}
\title{Exploring the $P$-wave bottom-strange molecular pentaquarks with the complex scaling method}
\author{Yuhang Zheng}\email{242238010@csu.edu.cn}
\author{Qing-Fu Song}\email{242201003@csu.edu.cn}
\author{Xiaonu Xiong}\email{xnxiong@csu.edu.cn}

\affiliation{School of Physics, Central South University, Changsha 410083, China}

\begin{abstract}
In this work, we perform a systematic investigation of the $P$-wave $\Sigma_b K/\Lambda_b K^{*}/\Sigma_b K^{*}$ and $\Sigma_b\bar{K}/\Lambda_b\bar{K}^{*}/\Sigma_b\bar{K}^{*}$ systems for all possible $I(J^P)$ combinations. In contrast to our earlier study of the negative-parity sector [Eur. Phys. J. C \textbf{85}, 1026 (2025)], the positive-parity sector is dominated by resonances. For the coupled-channel $I(J^P)=1/2(1/2^+)$ $\Sigma_b K/\Lambda_b K^*/\Sigma_b K^*$ system, we predict a narrow resonance above the $\Sigma_b K$ threshold. In the $I(J^P)=1/2(3/2^+)$ $\Lambda_b K^*/\Sigma_b K^*$ system, a molecular resonance candidate is found above the $\Lambda_b K^*$ threshold. For the higher-isospin $I=3/2$ sector, resonances are identified in the $\Sigma_b K^*$ subsystem for both $J^P=1/2^+$ and $3/2^+$ cases, though both states exhibit relatively compact sizes. We also extend our analysis to the $\Sigma_b\bar{K}/\Lambda_b\bar{K}^{*}/\Sigma_b\bar{K}^{*}$ systems, where a cutoff-sensitive resonance is found in the $I(J^P)=1/2(1/2^+)$ configuration. A broad molecular resonance candidate is found below the $\Sigma_b\bar{K}^*$ threshold in the $I(J^P)=1/2(3/2^+)$ system. We hope that these predictions provide useful guidance for future experimental searches for bottom-strange molecular pentaquarks by LHCb.

\end{abstract}


\maketitle
\section{Introduction}
Over the past two decades, a rapidly growing number of hadronic states have been discovered that cannot be naturally accommodated within the conventional constituent quark model classification of mesons and baryons. A prominent example is the X(3872), first observed by the Belle Collaboration in the $B^{\pm}\to K^{\pm}\pi^{+}\pi^{-}J/\psi$ decay~\cite{Belle:2003nnu}, which has stimulated extensive experimental and theoretical investigations~\cite{Lee:2009hy,Liu:2019zoy,Esposito:2025hlp}. Interestingly, many candidates for exotic states appear close to two-hadron thresholds, naturally suggesting a hadronic molecule interpretation: bound states or resonances of color-singlet hadrons held together by residual strong interactions~\cite{Dong:2017gaw,Guo:2017jvc}. Despite the appeal of the hadronic molecule interpretation for near-threshold resonances, alternative configurations such as compact multiquark states and hybrids remain viable candidates~\cite{Zou:2013af,Zou:2021sha,Karliner:2017qhf,Chen:2022asf}. Identifying their true nature demands a comprehensive analysis of the pole positions, production and decay patterns for each state.

Most previous investigations of hadronic molecules have focused on configurations in which the $S$ wave dominates, because the absence of a centrifugal barrier generally favors near-threshold binding. Orbital excitations, however, introduce qualitatively different dynamics into molecular systems. In a $P$-wave configuration, the centrifugal barrier tends to suppress the formation of bound states but, together with an attractive hadron-hadron interaction, may generate hadronic molecular states. Recent experimental and theoretical studies of near-threshold structures have stimulated renewed interest in $P$-wave molecular dynamics. In 2020, through an amplitude analysis of the $B^+\to D^+D^-K^+$ decay, the LHCb Collaboration reported two fully open-flavor structures in the $D^-K^+$ invariant mass distribution, denoted by $X_0(2900)$ and $X_1(2900)$~\cite{LHCb:2020bls,LHCb:2020pxc}. Their proximity to the $\bar D^*K^*$ threshold has motivated various interpretations in terms of $\bar D^{(*)}K^{(*)}$ hadronic molecules~\cite{Chen:2020aos,He:2020btl,Burns:2020epm,Agaev:2020nrc,Xiao:2020ltm,Ke:2022ocs}. In particular, a recent coupled-channel study proposed that the $X_1(2900)$ enhancement is associated with two poles dominated by $P$-wave $\bar D^*K^*$ configurations, denoted by $T_{cs1^-}(2900)$ and $T'_{cs1^-}(2900)$~\cite{Wang:2024ukc}. This unified $S$- and $P$-wave molecular picture of $X_0(2900)$ and $X_1(2900)$ encourages further searches for higher partial-wave states in open-charm systems. Recently, systematic calculations of the strong decay properties of $D_{s1}(2700)$ favor its assignment as a $P$-wave $DK^*$ molecular resonance~\cite{Lu:2026klm}. These developments suggest that orbitally excited hadronic molecules may constitute an experimentally accessible sector of exotic hadron spectroscopy.

Since the light diquark inside a singly heavy baryon shares the same color antitriplet $\bar{\mathbf{3}}_c$ representation as a light antiquark $\bar q$, a correspondence can be established between heavy mesons $Q\bar q$ and singly heavy baryons $Q[qq]_{\bar{\mathbf{3}}_c}$. This correspondence motivates the search for meson-baryon for the process molecular counterparts to the $D^{(*)}K^{(*)}$ molecular candidates~\cite{Chen:2023qlx,Sheng:2024hkf,Yan:2026ryi}. Ref.~\cite{Chen:2023qlx} investigated systems composed of charmed baryons $\Lambda_c/\Sigma_c$ and strange mesons $K^{(*)}$ using the one-boson-exchange (OBE) model, and proposed several bound states as possible meson-baryon molecular counterparts to the observed $T_{c\bar{s}0}^{++}(2900)$. Motivated further by heavy-flavor symmetry, we extended this investigation to the bottom sector by combining the OBE model with the complex scaling method (CSM) in our previous work~\cite{Song:2025yut}. In particular, a state with $I(J^P)=1/2(1/2^-)$ was obtained in the coupled $\Sigma_b\bar K/\Lambda_b\bar K^*/\Sigma_b\bar K^*$ system near the mass of the $\Xi_b(6227)$~\cite{LHCb:2018vuc}, suggesting a possible hadronic-molecular interpretation of this experimentally observed baryon. However, that study concentrated primarily on negative-parity configurations dominated by the $S$-wave, leaving the positive-parity molecular spectrum largely unexplored. In a recent work, several $P$-wave hidden-bottom molecular states were predicted using the OBE model~\cite{Wan:2026xzg}, which found that coupled-channel effects play an important role in their formation.

We use the OBE model combined with the CSM to systematically investigate positive-parity $P$-wave bottom-strange molecular pentaquarks in both the $Y_bK^{(*)}$ and $Y_b\bar K^{(*)}$ sectors, where $Y_b=\Lambda_b,\Sigma_b$. This framework has been successfully applied in our previous studies~\cite{Song:2025ijd,Su:2025toa,Song:2024ngu}. In contrast to the negative-parity states considered in our previous work~\cite{Song:2025yut}, which were predominantly generated by an attractive $S$-wave interaction, the positive-parity states studied here arise from a qualitatively different mechanism involving the interplay among the $P$-wave centrifugal barrier, attractive hadron-hadron interactions, and coupled-channel dynamics. In the $Y_b K^{(*)}$ sector, four resonance candidates are predicted: an $I(J^P)=1/2(1/2^+)$ state in the fully coupled $\Sigma_bK/\Lambda_bK^*/\Sigma_bK^*$ system; an $I(J^P)=1/2(3/2^+)$ state in the coupled $\Lambda_b K^* / \Sigma_b K^*$ system; and two $I=3/2$ states with $J^P=1/2^+$ and $3/2^+$ in the $\Sigma_bK^*$ system. In the $Y_b\bar K^{(*)}$ sector, the $I(J^P)=1/2(1/2^+)$ state exhibits a strong sensitivity to the cutoff parameter, and a resonance candidate is identified in the $I(J^P)=1/2(3/2^+)$ coupled $\Lambda_b\bar K^*/\Sigma_b\bar K^*$ system. These positive-parity resonances complement the negative-parity molecular spectrum predicted in our previous $S$-$D$-wave analysis and demonstrate that orbital excitation and coupled-channel effects can generate a distinct family of resonance poles in bottom-strange meson-baryon systems.

The rest of this paper is organized as follows. In Sec.~\ref{model}, we briefly introduce the formalism of effective interactions and review the complex scaling method. In Sec.~\ref{results}, we present the numerical results and discussion of the $\Sigma_bK/\Lambda_bK^{*}/\Sigma_bK^{*}$ and $\Sigma_b\bar K/\Lambda_b\bar K^*/\Sigma_b\bar K^*$ systems. Finally, we summarize our findings in Sec.~\ref{summary}. Numerical stability analysis and the explicit expressions of the scattering amplitudes are compiled in Appendices~\ref{App:further_discussion} and \ref{App:Scattering_amplitude}, respectively.

\section{Formalism of effective interactions and the complex scaling method}\label{model}
\subsection{Effective interactions}
The OBE model extends Yukawa’s meson theory by including the exchange of other light mesons such as $\sigma, \eta, \rho,$ and $\omega$~\cite{Liu:2018bkx,Machleidt:2017vls}. This framework allows us to systematically explore possible bound states and resonances in open-bottom meson-baryon systems. In this section, we employ the OBE model to derive the interaction potentials and investigate the formation of molecular states. Following the framework of Ref.~\cite{Liu:2011xc}, the effective Lagrangians with chiral symmetry that describe the interactions between the heavy baryons and the light mesons are constructed as
\bseq
\begin{align}
\mathcal{L}_{\mathcal{B}_{\bar{3}}} &= l_B \text{tr}[\bar{\mathcal{B}}_{\bar{3}}\sigma\mathcal{B}_{\bar{3}}]
+i\beta_B \text{tr}[\bar{\mathcal{B}}_{\bar{3}}v_{\alpha}
\left(\mathcal{V}^{\alpha}-\rho^{\alpha}\right)\mathcal{B}_{\bar{3}}],\label{lag1}\\
\mathcal{L}_{\mathcal{B}_{6}} &= l_S \text{tr}[\bar{\mathcal{S}}_{\mu}\sigma\mathcal{S}^{\mu}]
-\frac{3}{2}g_1\varepsilon^{\mu\nu\lambda\kappa}v_{\kappa}
\text{tr}[\bar{\mathcal{S}}_{\mu}A_{\nu}\mathcal{S}_{\lambda}]\nonumber\\
&\quad +i\beta_{S} \text{tr}[\bar{\mathcal{S}}_{\mu}v_{\alpha}
\left(\mathcal{V}^{\alpha}-\rho^{\alpha}\right) \mathcal{S}^{\mu}]
+\lambda_S \text{tr}[\bar{\mathcal{S}}_{\mu}F^{\mu\nu}\mathcal{S}_{\nu}],\\
\mathcal{L}_{\mathcal{B}_{\bar{3}}\mathcal{B}_6} &= ig_4 \text{tr}[\bar{\mathcal{S}}_{\mu}A^{\mu}\mathcal{B}_{\bar{3}}]
+i\lambda_I\varepsilon^{\mu\nu\lambda\kappa}v_{\mu} \text{tr}[\bar{\mathcal{S}}_{\nu}F_{\lambda\kappa}\mathcal{B}_{\bar{3}}]+ \text{h.c.}
\end{align}
\eseq
in which the spin-$1/2$ and spin-$3/2$ sextet baryons are combined into a superfield $\mathcal{S}_{\mu}$, defined as
\begin{equation}
\mathcal{S}_{\mu} = -\sqrt{\frac{1}{3}}(\gamma_\mu + v_\mu)\gamma^5 \mathcal{B}_6 + \mathcal{B}_{6\mu}^*,
\end{equation}
where $v_\mu=(1,\textbf{0})$ is the four-velocity of the heavy baryon. The matrices $\mathcal{B}_{\bar{3}}$ and $\mathcal{B}_6^{(*)}$ represent the ground-state multiplets of singly heavy baryons in the $\bar{3}_F$ and $6_F$ representations and are given by
\bseq
\begin{align}
\mathcal{B}_{\bar{3}} &= \begin{pmatrix}
0 & \Lambda_b^0 & \Xi_b^0 \\
-\Lambda_b^0 & 0 & \Xi_b^- \\
-\Xi_b^0 & -\Xi_b^- & 0
\end{pmatrix},\\
\quad
\mathcal{B}_6^{(*)} &= \begin{pmatrix}
\Sigma_b^{(*)+} & \frac{1}{\sqrt{2}}\Sigma_b^{(*)0} & \frac{1}{\sqrt{2}}\Xi_b^{(\prime, * )0} \\
\frac{1}{\sqrt{2}}\Sigma_b^{(*)0} & \Sigma_b^{(*)-} & \frac{1}{\sqrt{2}}\Xi_b^{(\prime, * )-} \\
\frac{1}{\sqrt{2}}\Xi_b^{(\prime, * )0} & \frac{1}{\sqrt{2}}\Xi_b^{(\prime, * )-} & \Omega_b^{(*)-}
\end{pmatrix}.\end{align}
\eseq
The axial current $A_{\mu}$, the vector current $\mathcal{V}_{\mu}$, and the vector meson field strength tensor $F_{\mu\nu}$ are defined as
\bseq
\begin{align}
\mathcal{V}_{\mu} &= \frac{1}{2}(\xi^{\dagger}\partial_{\mu}\xi+\xi\partial_{\mu}\xi^{\dagger}),\\
A_{\mu} &= \frac{1}{2}(\xi^{\dagger}\partial_{\mu}\xi-\xi\partial_{\mu}\xi^{\dagger}),\\
F_{\mu\nu} &= \partial_{\mu}\rho_{\nu}-\partial_{\nu}\rho_{\mu}+[\rho_{\mu},\rho_{\nu}].
\end{align}
\eseq
Here, $\xi=\exp(iP/f_{\pi})$ and $\rho_{\mu}=ig_V V_{\mu}/\sqrt{2}$, while $P$ and $V$ denote the light pseudoscalar and vector meson nonets, respectively. Their explicit forms are given by
\bseq
\begin{align}
P &= \begin{pmatrix}
\frac{\pi^0}{\sqrt{2}}+\frac{\eta}{\sqrt{6}} &\pi^+&K^+\\
\pi^-&-\frac{\pi^0}{\sqrt{2}}+\frac{\eta}{\sqrt{6}} &K^0\\
K^- &\bar{K}^0&-\frac{2}{\sqrt{6}}\eta
\end{pmatrix}, \\
V^{\mu} &= \begin{pmatrix}
\frac{\rho^0}{\sqrt{2}}+\frac{\omega}{\sqrt{2}}&\rho^+&K^{*+}\\
\rho^- &-\frac{\rho^0}{\sqrt{2}}+\frac{\omega}{\sqrt{2}} &K^{*0}\\
K^{*-}&\bar{K}^{*0}&\phi
\end{pmatrix}^{\mu}.\end{align}
\eseq
Following Ref.~\cite{Lin:1999ad}, the effective Lagrangians for the interactions among light mesons are given by
\bseq
\begin{align}
\mathcal{L}_{PPV} &= \frac{ig}{2\sqrt{2}} \text{tr}[\partial^{\mu}P (PV_{\mu}-V_{\mu}P)], \label{lag4}\\
\mathcal{L}_{VVP} &= \frac{g_{VVP}}{\sqrt{2}}\epsilon^{\mu\nu\alpha\beta} \text{tr}[\partial_{\mu}V_{\nu}\partial_{\alpha}V_{\beta}P],\label{lag5}\\
\mathcal{L}_{VVV} &= \frac{ig}{2\sqrt{2}} \text{tr}[\partial^{\mu}V^{\nu} (V_{\mu}V_{\nu}-V_{\nu}V_{\mu})]. \label{lag6}
\end{align}
\eseq
After substituting the superfield $\mathcal{S}_\mu$ and retaining only the leading terms for spin-1/2 components, we obtain the explicit effective Lagrangians describing the interactions between the heavy baryons and the light mesons as
\bseq
\begin{align}
\mathcal{L}_{\sigma} =& l_B \text{tr}[\bar{\mathcal{B}}_{\bar{3}}\sigma\mathcal{B}_{\bar{3}}] - l_S \text{tr}[\bar{\mathcal{B}}_6\sigma\mathcal{B}_6],\\
\mathcal{L}_{P} =& i\frac{g_1}{2f_{\pi}}\varepsilon^{\mu\nu\lambda\kappa}v_{\kappa} \text{tr}[\bar{\mathcal{B}}_6 \gamma_{\mu}\gamma_{\lambda}\partial_{\nu}{P}\mathcal{B}_6] \nonumber\\
&- \frac{g_4}{\sqrt{3}f_{\pi}} \text{tr}[\bar{\mathcal{B}}_6\gamma^5 \left(\gamma_{\mu}+v_{\mu}\right)\partial^{\mu}{P}\mathcal{B}_{\bar{3}}] + \text{h.c.},\\
\mathcal{L}_{V} =& \frac{\beta_Bg_V }{\sqrt{2}}\text{tr}[\bar{\mathcal{B}}_{\bar{3}}v_{\mu}V^{\mu}\mathcal{B}_{\bar{3}}] -\frac{\beta_Sg_V}{\sqrt{2}} \text{tr}[\bar{\mathcal{B}}_6v_{\mu}V^{\mu}\mathcal{B}_6] \nonumber\\
&- \frac{\lambda_Ig_V}{\sqrt{6}}\varepsilon^{\mu\nu\lambda\kappa}v_{\mu} \text{tr}[\bar{\mathcal{B}}_6\gamma^5\gamma_{\nu} \left(\partial_{\lambda} {V}_{\kappa}-\partial_{\kappa} {V}_{\lambda}\right)\mathcal{B}_{\bar{3}}] \nonumber\\
&- i\frac{\lambda_S g_V}{3\sqrt{2}} \text{tr}[\bar{\mathcal{B}}_6\gamma_{\mu}\gamma_{\nu} \left(\partial^{\mu} {V}^{\nu}-\partial^{\nu} {V}^{\mu}\right) \mathcal{B}_6] + \text{h.c.}.
\end{align}
\eseq
Additionally, the effective Lagrangians describing the interactions between the strange mesons and the light mesons are given by
\bseq
\begin{align}
\mathcal{L}_{K^{(*)}K^{(*)}\sigma} =& g_{\sigma }m_K\bar{K} K\sigma - g_{\sigma }m_{K^*}\bar{K}^{*}\cdot K^{*}\sigma,\\
\mathcal{L}_{P KK^*} =& \frac{ig}{4}\bigg[ \left(\bar{K}^{*\mu} K - \bar{K} K^{*\mu}\right)\left(\bm{\tau}\cdot\partial_{\mu}\bm{\pi}+\frac{\partial_{\mu}{\eta}}{\sqrt{3}}\right)\nonumber\\
& + \left(\partial_{\mu}\bar{K} K^{*\mu} - \bar{K}^{*\mu}\partial_{\mu}K\right)\left(\bm{\tau}\cdot\bm{\pi}+\frac{\eta}{\sqrt{3}}\right) \bigg],\\
\mathcal{L}_{{V} KK} =& \frac{ig}{4}\left[\bar{K}\partial_{\mu}K - \partial_{\mu}\bar{K}K\right]\left(\bm{\tau}\cdot\bm{\rho}^{\mu}+{\omega}^{\mu}\right),\\
\mathcal{L}_{{V} K^*K^*} =& \frac{ig}{4} \bigg[ \left(\bar{K}_{\mu}^*\partial^{\mu}K^{*\nu}-\partial^{\mu}\bar{K}^{*\nu} K_{\mu}^*\right)\left(\bm{\tau}\cdot\bm{\rho}_{\nu}+\omega_{\nu}\right)\nonumber\\
&+ \left(\partial^{\mu}\bar{K}^{*\nu}K_{\nu}^*-\bar{K}_{\nu}^*\partial^{\mu}K^{*\nu}\right) \left(\bm{\tau}\cdot\bm{\rho}_{\mu}+\omega_{\mu}\right)\nonumber\\
&+ \left(\bar{K}_{\nu}^* K^*_{\mu}-\bar{K}_{\mu}^*K^*_{\nu}\right) \left(\bm{\tau}\cdot\partial^{\mu}\bm{\rho}^{\nu}+\partial^{\mu}\omega^{\nu}\right) \bigg],\\
\mathcal{L}_{P K^*K^*} =& g_{VVP}\varepsilon_{\mu\nu\alpha\beta} \partial^{\mu}\bar{K}^{*\nu}\partial^{\alpha}K^{*\beta}\left(\bm{\tau}\cdot\bm{\pi}+\frac{\eta}{\sqrt{3}}\right),\\
\mathcal{L}_{V KK^*} = & g_{VVP}\varepsilon_{\mu\nu\alpha\beta} \left(\partial^{\mu}\bar{K}^{*\nu}K + \bar{K}\partial^{\mu}{K}^{*\nu}\right) \nonumber \\
& \times \left(\bm{\tau}\cdot\partial^{\alpha}\bm{\rho}^{\beta}+\partial^{\alpha}{\omega}^{\beta}\right).
\end{align}
\eseq
The values of the relevant parameters in the Lagrangians are listed in Table~\ref{total_info}~\cite{Liu:2011xc,Chen:2017xat,Kaymakcalan:1983qq,ParticleDataGroup:2026mpi}.
\begin{table}[htbp] 
\centering 
\begin{ruledtabular}
\begin{tabular}{lc|lc} 
\textbf{Parameter}& \textbf{Value}& \textbf{Hadron} & \textbf{Mass (GeV)} \\ \hline
$l_S=-2l_B$& 7.3& ${\pi}$ & 0.13957 \\
$g_1=(\sqrt{8}/3)g_4$ & 1.0& ${\eta}$ & 0.54786 \\
$\beta_S g_V=-2\beta_Bg_V$& 12.0 & ${\rho}$ & 0.77526 \\
$\lambda_Sg_V=-2\sqrt{2}\lambda_Ig_V$& $19.2~\text{GeV}^{-1}$ & ${\omega}$ & 0.78266 \\
$g_\sigma$ & $-3.65$ & ${\sigma}$ & 0.60000 \\ 
$g$ & 12.00& ${K}$ & 0.49368 \\
$g_{VVP}$ & $7.33~\text{GeV}^{-1}$ & ${K^{*}}$& 0.89188 \\
$f_{\pi}$ & $0.132~\text{GeV}$& ${\Lambda_b}$ & 5.61957 \\ 
 & & ${\Sigma_b}$ & 5.81056 \\
\end{tabular}
\end{ruledtabular}
\caption{\label{total_info} The relevant parameters and hadron masses adopted in this work.}
\end{table}

To obtain the interaction potential in coordinate space, we first apply the Breit approximation, which reduces the relativistic scattering amplitude to an instantaneous effective potential in momentum space. This approximation is valid in the low-energy regime characterized by small momentum transfer, enabling the extraction of various interaction terms, including the central, spin-spin, tensor, and spin-orbit components~\cite{Breit:1929zz,Breit:1930zza}. For a specified meson-baryon scattering process $h_{1}h_{2}\to h_{3}h_{4}$, the effective potential in momentum space is defined as
\begin{equation}\label{breit}
\mathcal{V}_j^{h_{1}h_{2}\to h_{3}h_{4}}(\bm{q}) = -\frac{\mathcal{M}_j(h_{1}h_{2}\to h_{3}h_{4})}{4\sqrt{m_{1}m_{2}m_{3}m_{4}}},
\end{equation}
where $\mathcal{M}_j$ denotes the scattering amplitude associated with the exchange of meson $j$. Here, $m_1$ and $m_3$ are the masses of the baryons $h_1$ and $h_3$, while $m_2$ and $m_4$ are the masses of the mesons $h_2$ and $h_4$, respectively. The corresponding potential in coordinate space is then obtained by a Fourier transformation
\begin{equation}
\mathcal{V}_j^{h_{1}h_{2}\to h_{3}h_{4}}(\bm{r})=\int\frac{d^3\bm{q}}{(2\pi)^3}e^{i\bm{q}\cdot\bm{r}}\mathcal{V}_j^{h_{1}h_{2}\to h_{3}h_{4}}(\bm{q})\mathcal{F}^2(\bm{q}^2,m_j^2),
\end{equation}
where $m_j$ is the mass of the exchanged meson $j$, and $\mathcal{F}$ denotes the form factor given by~\cite{Machleidt:1987hj,Li:2012bt}
\begin{equation}\label{FF}
\mathcal{F}(\bm{q}^2,m_j^2)=\frac{\Lambda^2-m_j^2}{\Lambda^2+\bm{q}^2}.
\end{equation}
Here, $\Lambda$ is a UV cutoff, which accounts for the intrinsic finite size of the interacting hadrons.

To incorporate the energy transfer $q^0$ in the OBE potential, we define the effective mass $\tilde{m}_j$ and the effective cutoff $\tilde{\Lambda}$ for the exchanged meson $j$ as
\begin{equation}\label{eq:m_eff}
\tilde{m}_j =\sqrt{\left|m_j^2-(q^0)^2\right|},\qquad\tilde{\Lambda} =\sqrt{\Lambda^2-(q^0)^2},
\end{equation}
where $q^0$ is given by
\begin{equation}
q^0 =\frac{m_1^2+m_4^2-m_2^2-m_3^2}{2(m_3+m_4)},
\end{equation}
for the process $h_{1}h_{2}\to h_{3}h_{4}$. Following the framework described above, the explicit forms of all scattering amplitudes for the $\Sigma_bK/\Lambda_bK^{*}/\Sigma_bK^{*}$ systems are provided in Appendix~\ref{App:Scattering_amplitude}.

\subsection{Complex scaling method}\label{sec3}

In this subsection, we briefly review the complex scaling method~\cite{Aguilar:1971ve,Balslev:1971vb,Simon:1972qft}, which is employed to solve the coupled-channel Schr\"odinger equation and to identify resonance poles. Resonances correspond to poles of the scattering matrix and are characterized by complex energies $E_{\rm pole}=E-i\Gamma/2$, where $E$ and $\Gamma$ denote the resonance energy and decay width, respectively. In the CSM, the resonance wave function after complex scaling becomes square-integrable, permitting its treatment in a manner analogous to
that of bound states. This method was originally developed for the study of atomic and molecular systems~\cite{Reinhardt:1982com,Ho:1983lwa,Moiseyev:1998gjp} and has since been widely applied to resonances in nuclear and hadronic systems~\cite{Myo:2014ypa, Yu:2021lmb}.

Under the complex scaling transformation, the relative coordinate $\bm r$ and the conjugate momentum $\bm p$ are transformed by the scaling operator $U(\theta) = e^{-\theta U_D}$ with the generator being $U_D = \frac{1}{2}(\bm r \cdot \bm p + \bm p \cdot \bm r)$~\cite{Aguilar:1971ve} according to
\begin{equation}
\begin{split}
\bm r^{\,\theta}&=U(\theta)\bm r U^{-1}(\theta)=\bm r e^{i\theta},\\
\bm p^{\,\theta}&=U(\theta)\bm p U^{-1}(\theta)=\bm p e^{-i\theta},
\end{split}
\end{equation}
where $\theta>0$ is the scaling angle. The Schr\"odinger equation then transforms as
\begin{equation}
H^\theta \Psi^\theta
=
E^\theta \Psi^\theta,
\label{eq:csm-sch}
\end{equation}
in which $H^\theta = U(\theta)HU^{-1}(\theta)$ and $E^\theta$ denote the scaled Hamiltonian and the scaled energy, respectively.

\begin{figure}[htbp]
\centering
\includegraphics[width=0.8\linewidth]{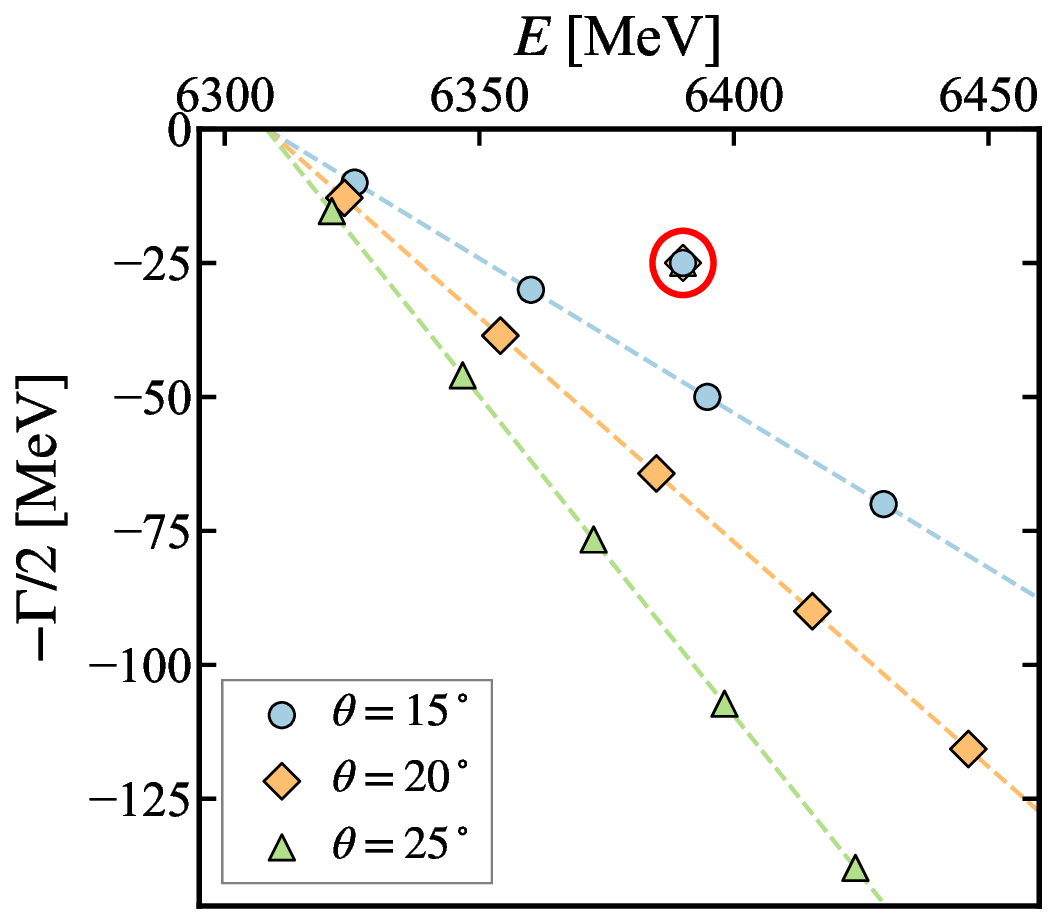}
\caption{
Distribution of the eigenvalues of the complex-scaled Hamiltonian $H^\theta$. The encircled eigenvalue denotes the resonance identified by the CSM.}
\label{csmp}
\end{figure}

In the CSM, the continuum states are rotated into the complex-energy plane by an angle of $-2\theta$ relative to the real axis. In contrast, bound-state poles remain on the real axis, while resonance poles are separated from the rotated continuum states; both bound-state and resonance poles are independent of $\theta$. A resonance is accordingly identified by both its separation from the rotated continuum trajectories and its stability against variations in the scaling angle $\theta$. A schematic distribution of the complex-scaled eigenvalues is shown in Fig.~\ref{csmp}.

For a coupled-channel system, the complex-scaled radial Schr\"odinger equation is given by
\begin{equation}
\sum_k\left[\left(T_j e^{-2i\theta}+W_j-E^\theta\right)\delta_{jk}+V_{jk}(re^{i\theta})\right]u_k^\theta(r)=0,
\label{Seq}
\end{equation}
where $j$ and $k$ label the coupled channels, $u_k^\theta(r)$ denotes the reduced radial wave function, $W_j$ is the corresponding mass threshold of the constituent hadrons in channel $j$, and $V_{jk}$ is the interaction potential. Specifically, $V_{jj}$ denotes the diagonal potential for channel $j$, while $V_{jk}$ ($j \neq k$) represents the off-diagonal potential between channels $j$ and $k$. The radial kinetic energy operator $T_{j}$ is defined as
\begin{equation}
T_{j} = \frac{1}{2\mu_j} \left( -\frac{d^2}{dr^2} + \frac{l_j(l_j+1)}{r^2} \right),
\end{equation}
where $l_j$ and $\mu_j$ denote the orbital angular momentum quantum number and the reduced mass of channel $j$, respectively. 
To solve Eq.~(\ref{Seq}) numerically, the reduced radial wave function in each channel is expanded in a finite harmonic oscillator (HO) basis 
\begin{equation}
u_k^\theta(r)=\sum_{n=0}^{N-1}c_{kn}^\theta\,u_{n l_k}(r),
\label{eq:ho-expansion}
\end{equation}
where $N$ denotes the number of basis functions. In practice, we find that $N\approx 250$ already gives satisfactorily converged results, as demonstrated in Appendix~\ref{App:further_discussion}. To be conservative, we adopt $N = 350$ for all calculations in this work. The explicit form of the HO basis reads~\cite{Yamani:1975jmm}
\begin{equation}
u_{n l_k}(r)=\sqrt{\frac{2n!}{b_0\Gamma(n+l_k+3/2)}}\left(\frac{r}{b_0}\right)^{l_k+1}e^{-r^2/(2b_0^2)}L_n^{(l_k+1/2)}\left(\frac{r^2}{b_0^2}\right),
\end{equation}
where $b_0$ is the oscillator length, $l_k$ is the orbital angular momentum quantum number for channel $k$, and $L_n^{(l_k+1/2)}$ denotes the generalized Laguerre polynomials. Although observables calculated with a complete HO basis are independent of the oscillator length $b_0$, those calculated with a truncated HO basis generally depend on $b_0$, necessitating the identification of a stability range where the results are insensitive to $b_0$~\cite{Wang:2021hho,Sharaf:2019jgw,Caprio:2012as}. As exemplified by Fig.~\ref{fig:stability_analysis}, the real part of the RMS radius $\operatorname{Re}(r_{\text{RMS}})$, the energy $E$, and the decay width $\Gamma$ remain stable in $b_0 \in [2.4, 4.0]~\mathrm{GeV}^{-1}$. We adopt $b_0=3.0~\mathrm{GeV}^{-1}$ throughout this paper.

Based on the complex-scaled wave functions, quantities such as the root-mean-square (RMS) radius $r_{\text{RMS}}$ and the channel proportions $P_i$ for resonances can be evaluated using the c-product~\cite{Homma:1997wtc, Lin:2023ihj, Moiseyev:2011}
\begin{align}
r_{\text{RMS}}^2&=(\Psi^{\theta}|r^2|\Psi^{\theta})=\sum_i\int_0^\infty dr\,u_i^\theta(r)\,r^2\,u_i^\theta(r),
\label{eq:rms-cproduct}\\
P_i&=(\Psi_i^\theta|\Psi_i^\theta)=\int_0^\infty dr\,u_i^\theta(r)u_i^\theta(r),
\label{eq:channel-cproduct}
\end{align}
where $\Psi^{\theta}_i$ is the wave function of channel $i$. These wave functions satisfy the normalization condition
\begin{equation}
\sum_i P_i = 1. 
\end{equation}
Unlike the conventional Hermitian inner product, the c-product is defined by integrating the product of two wave functions without taking the complex conjugation. For resonances, $P_i$ are generally complex and do not represent standard probabilities~\cite{Berggren:1970wto}. Nevertheless, they can be used to identify the dominant components of the resonance. The real part of a complex radius characterizes the spatial size, whereas a sizable imaginary part reflects the resonant nature of the state~\cite{Gyarmati:1972yac,Homma:1997wtc,Myo:2023heu}. In the following discussion, states with $\operatorname{Re}(r_{\text{RMS}})<1$ fm are regarded as relatively compact and are therefore considered less favorable for a typical hadronic-molecular interpretation. Besides the RMS radius, threshold behavior, channel proportions, and cutoff dependence must also be taken into consideration when evaluating a state as a hadronic molecular candidate. 

To analyze the influence of specific channel couplings on a resonance pole, we employ the Feshbach projection formalism~\cite{Feshbach:1958nx,Feshbach:1962ut}. The channel space is partitioned into a primary subspace $P$ and its complementary subspace $Q$, such that the complex-scaled Hamiltonian $H^\theta$ is written in the block-matrix form
\begin{equation}
H^\theta = \begin{pmatrix} H_{PP}^\theta & H_{PQ}^\theta \\ H_{QP}^\theta & H_{QQ}^\theta \end{pmatrix}.
\end{equation}
According to Eq.~(\ref{Seq}), the radial kinetic energy operator $T_j$ and the threshold $W_j$ are diagonal in channel space. Therefore, the off-diagonal blocks $H_{PQ}^\theta$ and $H_{QP}^\theta$ contain only the off-diagonal potentials $V_{jk}(re^{i\theta})$, with $j\in P$ and $k\in Q$. Eliminating the $Q$-space component yields an effective equation in the $P$ space
\begin{equation}
\left[H_{PP}^\theta+\Sigma_P^\theta(E)\right]\Psi_P^\theta=E\Psi_P^\theta,
\end{equation}
where the self-energy operator is defined as~\cite{Saito:1994iz}
\begin{equation}
\Sigma_P^\theta(E)=H_{PQ}^\theta\left(E-H_{QQ}^\theta\right)^{-1}H_{QP}^\theta .
\end{equation}
To quantify the coupling effects at the resonance position, we evaluate the expectation value of the self-energy at the pole energy $E_{\rm pole}$
\begin{equation}
\bar{\Sigma}_P^\theta =\frac{(\Psi_P^\theta |\Sigma_P^\theta(E_{\rm pole}) |\Psi_P^\theta)}{(\Psi_P^\theta|\Psi_P^\theta)} ,
\end{equation}
where $\Psi_P^\theta$ denotes the $P$-space component of the total resonance wave function. To analyze these coupling effects in detail, we consider a two-channel coupling in which the $P$- and $Q$-spaces each contain a single channel, denoted by $j \in P$ and $k \in Q$, respectively. In this case, the self-energy can equivalently be written as
\begin{equation}
\bar{\Sigma}_P^\theta =\frac{(\Psi_j^\theta|V_{jk}(re^{i\theta})|\Psi_k^\theta)}{
(\Psi_j^\theta|\Psi_j^\theta)}.
\end{equation}
Thus, in this two-channel case, the self-energy is evaluated from the off-diagonal potential and the corresponding channel wave functions. A negative value of $\operatorname{Re}(\bar{\Sigma}_P^\theta)$ indicates that the coupling between these two channels yields a negative contribution to the pole energy and thereby manifests as an effective attraction induced by coupled-channel effects~\cite{Yamaguchi:2017zmn}. Consequently, this formalism provides a quantitative tool to evaluate the role of specific channel couplings in the formation of resonances.

\section{Results and discussion}\label{results}

\begin{table*}[htbp]
\centering
\renewcommand{\arraystretch}{1.6}
\setlength{\tabcolsep}{6pt}
\footnotesize

\begin{tabular}{l|ccc|lllll}
\hline\hline
\multirow{2}{*}{System} & \multirow{2}{*}{$\Lambda~[\mev]$} & \multirow{2}{*}{$E-i\Gamma/2~[\mev]$} & \multirow{2}{*}{$r_{\text{RMS}}~[\fm]$} & \multicolumn{5}{c}{$P(\%)$} \\ \cline{5-9}
& & & & $\Sigma_b K(^2P_{1/2})$ & $\Lambda_b K^*(^2P_{1/2})$ & $\Lambda_b K^*(^4P_{1/2})$ & $\Sigma_b K^*(^2P_{1/2})$ & $\Sigma_b K^*(^4P_{1/2})$ \\ \hline

\multirow{3}{*}{\begin{tabular}[c]{@{}l@{}} $1/2(1/2^+)$ \\ $\Sigma_b K / \Lambda_b K^* / \Sigma_b K^*$ \end{tabular}}
& 1515 & $6311.19-0.51i$ & $1.00+0.70i$ & $\mathbf{52.35+6.72i}$ & $2.33-0.33i$ & $28.40-3.99i$ & $2.44-0.35i$ & $14.48-2.05i$ \\
& 1520 & $6308.55-0.21i$ & $1.14+0.58i$ & $\mathbf{51.31+4.14i}$ & $2.37-0.20i$ & $28.95-2.45i$ & $2.52-0.22i$ & $14.85-1.27i$ \\
& 1525 & $6305.81-0.04i$ & $1.22+0.40i$ & $\mathbf{49.48+1.89i}$ & $2.45-0.09i$ & $29.96-1.12i$ & $2.64-0.10i$ & $15.46-0.58i$ \\ \hline

System & $\Lambda~[\mev]$ & $E-i\Gamma/2~[\mev]$& $r_{\text{RMS}}~[\fm]$ & $\Lambda_b K^*(^2P_{3/2})$ & $\Lambda_b K^*(^4P_{3/2})$ & $\Sigma_b K^*(^2P_{3/2})$ & $\Sigma_b K^*(^4P_{3/2})$ & \\ \hline
\multirow{3}{*}{\begin{tabular}[c]{@{}l@{}} $1/2(3/2^+)$ \\ $\Lambda_b K^* / \Sigma_b K^*$ \end{tabular}}
& 2340 & $6515.35-0.32i$ & $0.84+0.87i$ & $8.41+0.32i$ & $\mathbf{78.44+2.18i}$ & $0.82-0.15i$ & $12.34-2.34i$ & \\
& 2345 & $6514.20-0.12i$ & $1.03+1.09i$ & $8.36+0.24i$ & $\mathbf{78.37+1.60i}$ & $0.82-0.11i$ & $12.45-1.72i$ & \\
& 2350 & $6513.03-0.01i$ & $1.76+1.03i$ & $8.31+0.08i$ & $\mathbf{78.10+0.59i}$ & $0.83-0.04i$ & $12.77-0.63i$ & \\\hline

System & $\Lambda~[\mev]$ & $E-i\Gamma/2~[\mev]$& $r_{\text{RMS}}~[\fm]$ & $\Sigma_b K^*(^2P_{1/2})$ & $\Sigma_b K^*(^4P_{1/2})$ & & & \\ \hline
\multirow{3}{*}{\begin{tabular}[c]{@{}l@{}} $3/2(1/2^+)$ \\ $\Sigma_b K^*$ \end{tabular}}
& 1820 & $6709.10-0.70i$ & $0.70+0.71i$ & $\mathbf{97.95-1.04i}$ & $2.05+1.04i$ & & & \\
& 1825 & $6707.14-0.30i$ & $0.79+0.80i$ & $\mathbf{97.98-0.73i}$ & $2.02+0.73i$ & & & \\
& 1830 & $6705.12-0.03i$ & $1.14+1.10i$ & $\mathbf{98.06-0.32i}$ & $1.94+0.32i$ & & & \\\hline

System & $\Lambda~[\mev]$ & $E-i\Gamma/2~[\mev]$& $r_{\text{RMS}}~[\fm]$ & $\Sigma_b K^*(^2P_{3/2})$ & $\Sigma_b K^*(^4P_{3/2})$ & & & \\ \hline
\multirow{3}{*}{\begin{tabular}[c]{@{}l@{}} $3/2(3/2^+)$ \\ $\Sigma_b K^*$ \end{tabular}}
& 1820 & $6710.02-0.85i$ & $0.66+0.65i$ & $\mathbf{99.47-0.46i}$ & $0.53+0.46i$ & & & \\
& 1825 & $6707.97-0.42i$ & $0.73+0.71i$ & $\mathbf{99.46-0.31i}$ & $0.54+0.31i$ & & & \\
& 1830 & $6705.87-0.10i$ & $0.90+0.89i$ & $\mathbf{99.48-0.16i}$ & $0.52+0.16i$ & & & \\\hline\hline

System & $\Lambda~[\mev]$ & $E-i\Gamma/2~[\mev]$& $r_{\text{RMS}}~[\fm]$ & $\Sigma_b \bar{K}(^2P_{1/2})$ & $\Lambda_b \bar{K}^*(^2P_{1/2})$ & $\Lambda_b \bar{K}^*(^4P_{1/2})$ & $\Sigma_b \bar{K}^*(^2P_{1/2})$ & $\Sigma_b \bar{K}^*(^4P_{1/2})$ \\ \hline
\multirow{3}{*}{\begin{tabular}[c]{@{}l@{}} $1/2(1/2^+)$ \\ $\Sigma_b \bar{K}/\Lambda_b \bar{K}^*/\Sigma_b \bar{K}^*$ \end{tabular}}
& 2015 & $6314.67-1.10i$ & $0.74+0.78i$ & $\mathbf{74.05+5.36i}$ & $2.45-0.52i$ & $0.78-0.14i$ & $2.40-0.51i$ & $20.32-4.20i$ \\
& 2020 & $6310.42-0.39i$ & $1.01+0.68i$ & $\mathbf{73.69+3.20i}$ & $2.53-0.31i$ & $0.74-0.08i$ & $2.48-0.31i$ & $20.57-2.50i$ \\
& 2025 & $6305.93-0.04i$ & $1.17+0.42i$ & $\mathbf{72.21+1.06i}$ & $2.72-0.10i$ & $0.74-0.02i$ & $2.66-0.10i$ & $21.67-0.84i$ \\ \hline

System & $\Lambda~[\mev]$ & $E-i\Gamma/2~[\mev]$& $r_{\text{RMS}}~[\fm]$ & $\Lambda_b \bar{K}^*(^2P_{3/2})$ & $\Lambda_b \bar{K}^*(^4P_{3/2})$ & $\Sigma_b \bar{K}^*(^2P_{3/2})$ & $\Sigma_b \bar{K}^*(^4P_{3/2})$ & \\ \hline
\multirow{3}{*}{\begin{tabular}[c]{@{}l@{}} $1/2(3/2^+)$ \\ $\Lambda_b \bar{K}^* / \Sigma_b \bar{K}^*$ \end{tabular}}
& 2075 & $6701.06-8.84i$ & $1.05-0.33i$ & $0.06+2.01i$ & $-2.43-2.49i$ & $12.11-0.17i$ & $\mathbf{90.26+0.64i}$ & \\
& 2080 & $6699.35-9.03i$ & $1.03-0.28i$ & $0.15+2.11i$ & $-2.50-2.38i$ & $12.10-0.21i$ & $\mathbf{90.25+0.48i}$ & \\
& 2085 & $6697.55-9.19i$ & $1.01-0.24i$ & $0.24+2.20i$ & $-2.55-2.27i$ & $12.09-0.25i$ & $\mathbf{90.23+0.32i}$ & \\ \hline

System & $\Lambda~[\mev]$ & $E-i\Gamma/2~[\mev]$& $r_{\text{RMS}}~[\fm]$ & $\Sigma_b \bar{K}^*(^2P_{3/2})$ & $\Sigma_b \bar{K}^*(^4P_{3/2})$ & & & \\ \hline
\multirow{3}{*}{\begin{tabular}[c]{@{}l@{}} $1/2(3/2^+)$ \\ $\Sigma_b \bar{K}^*$ \end{tabular}}
& 2130 & $6706.70-0.55i$ & $1.12+1.32i$ & $9.36-0.46i$ & $\mathbf{90.64+0.46i}$ & & & \\
& 2135 & $6706.04-0.33i$ & $1.24+1.41i$ & $9.32-0.35i$ & $\mathbf{90.68+0.35i}$ & & & \\
& 2140 & $6705.35-0.14i$ & $1.47+1.59i$ & $9.29-0.23i$ & $\mathbf{90.71+0.23i}$ & & & \\\hline \hline
\end{tabular}
\caption{Numerical results for positive-parity $Y_bK^{(*)}$ and $Y_b\bar K^{(*)}$ systems. Proportions (\%) denote channel contributions, with dominant components in bold. Complex-valued entries correspond to resonance quantities. For each system, three representative resonances near the threshold are presented at 5~MeV intervals of the cutoff $\Lambda$.}
\label{tab:numerical_results}
\end{table*}

\begin{figure*}[htb!] 
\centering
\includegraphics[width=1\textwidth]{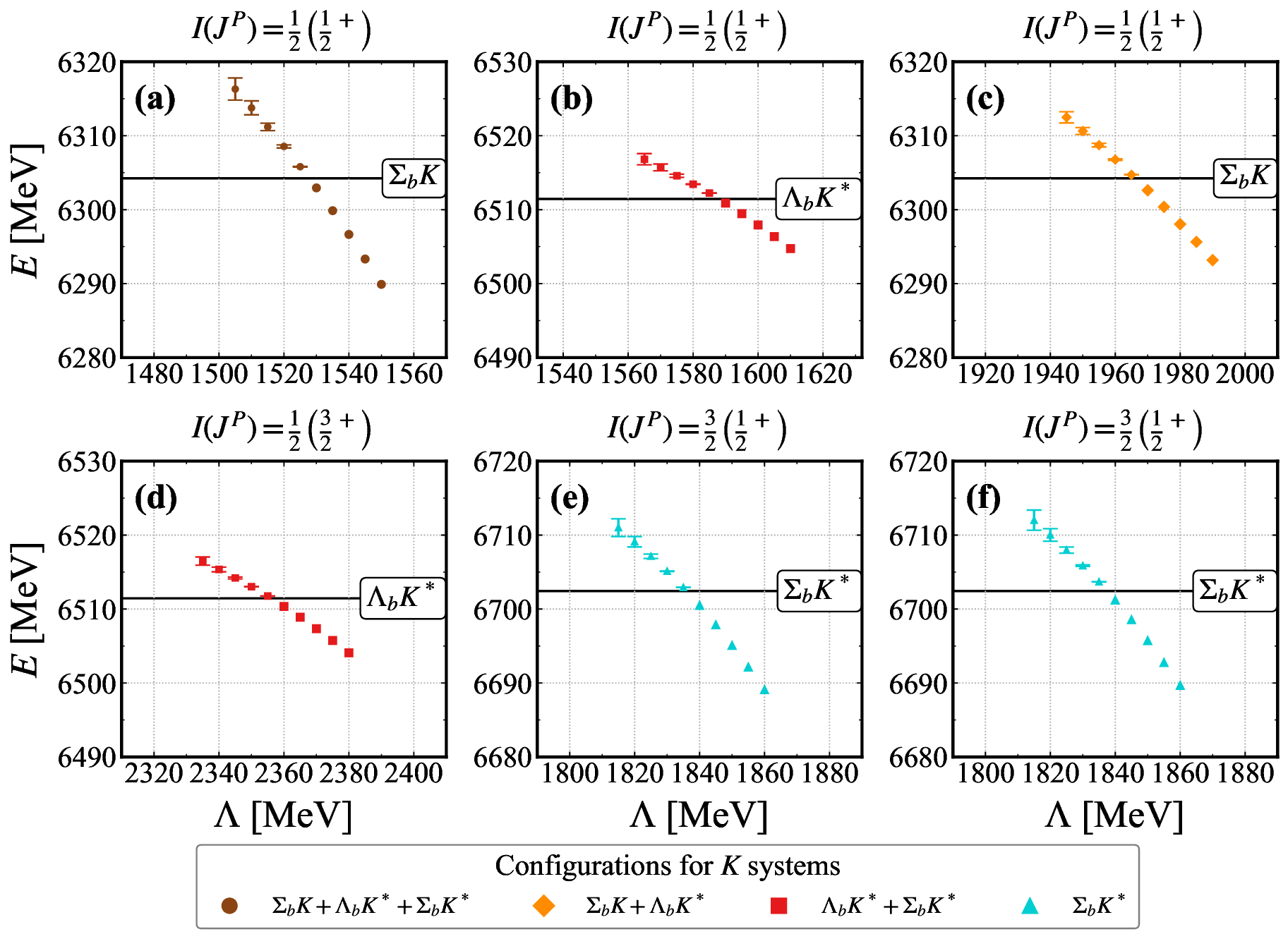}
\includegraphics[width=0.7\textwidth]{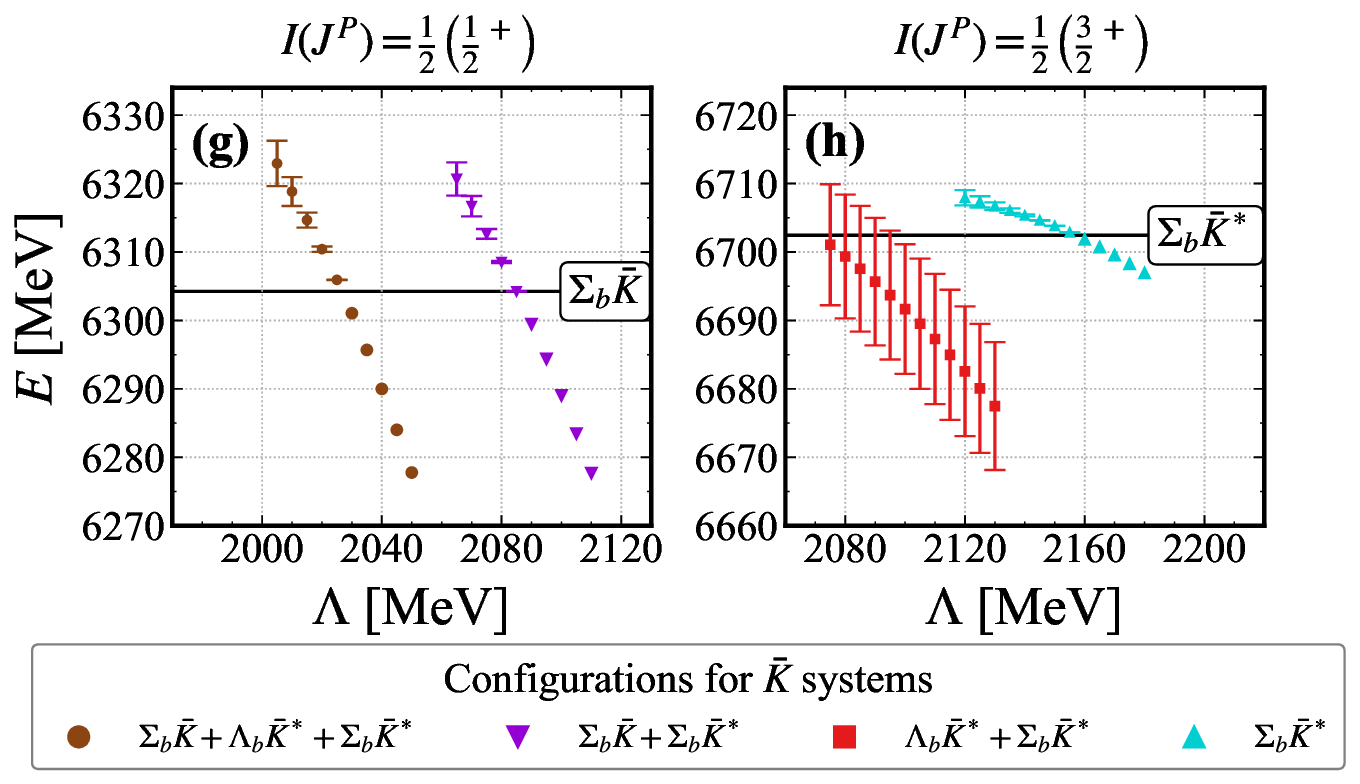}
\caption{The dependence of the complex energy eigenvalues on the cutoff $\Lambda$ for the $\Sigma_bK/\Lambda_bK^{*}/\Sigma_bK^{*}$ systems with $I(J^P) = 1/2(1/2^+,3/2^+)$, the $\Sigma_b K/\Sigma_b K^{*}$ systems with $I(J^P) = 3/2(1/2^+,3/2^+)$, and $\Sigma_b\bar K/\Lambda_b\bar K^*/\Sigma_b\bar K^*$ systems with $I(J^P) = 1/2(1/2^+,3/2^+)$. The horizontal axis represents the energy $E$, and the vertical error bars indicate the decay widths $\Gamma$.}
\label{fig:Energy_Spectrum_K}
\end{figure*}

By solving the complex-scaled radial Schrödinger equation through the HO basis expansion method, we identify several states in the positive-parity $\Sigma_b\bar{K}/\Lambda_b\bar{K}^{*}/\Sigma_b\bar{K}^{*}$ and $\Sigma_bK/\Lambda_bK^{*}/\Sigma_bK^{*}$ systems. In this model, the only adjustable parameter is the cutoff $\Lambda$, which accounts for the intrinsic finite size of the interacting hadrons. Although a cutoff of approximately $1000~\mathrm{MeV}$ has been successfully employed for the deuteron~\cite{Tornqvist:1993ng, Tornqvist:1993vu}, its value may extend up to around $2200~\mathrm{MeV}$ for hadron systems~\cite{Yang:2011wz}. In this study, we vary $\Lambda$ over the range of $1000$--$3000~\mathrm{MeV}$, following our previous work~\cite{Song:2025yut}, to systematically search for possible $P$-wave bound states and resonances, and to investigate their cutoff dependence.

The investigated configurations are categorized by their isospin and total angular momentum: the $\Sigma_bK/\Lambda_bK^{*}/\Sigma_bK^{*}$ systems with $I(J^P) = 1/2(1/2^+, 3/2^+)$, the $\Lambda_b K^{*}/\Sigma_b K^{*}$ systems with $I(J^P) = 1/2(5/2^+)$, the $\Sigma_b K/\Sigma_b K^{*}$ systems with $I(J^P) = 3/2(1/2^+, 3/2^+)$, and the $\Sigma_b K^{*}$ systems with $I(J^P) = 3/2(5/2^+)$. An analogous classification applies to the corresponding $\Sigma_b\bar{K}/\Lambda_b\bar{K}^{*}/\Sigma_b\bar{K}^{*}$ systems. Notably, for all configurations with $J^P = 5/2^+$, no bound states or resonances are identified within the scanned cutoff range, and they are thus excluded from further analysis. The numerical results are summarized in Fig.~\ref{fig:Energy_Spectrum_K} and Table~\ref{tab:numerical_results}.

\subsection{$\Lambda_b/\Sigma_b$ and $K^{(*)}$ interactions}
\begin{figure*}[htb!] 
\centering
\includegraphics[width=1.5\columnwidth]{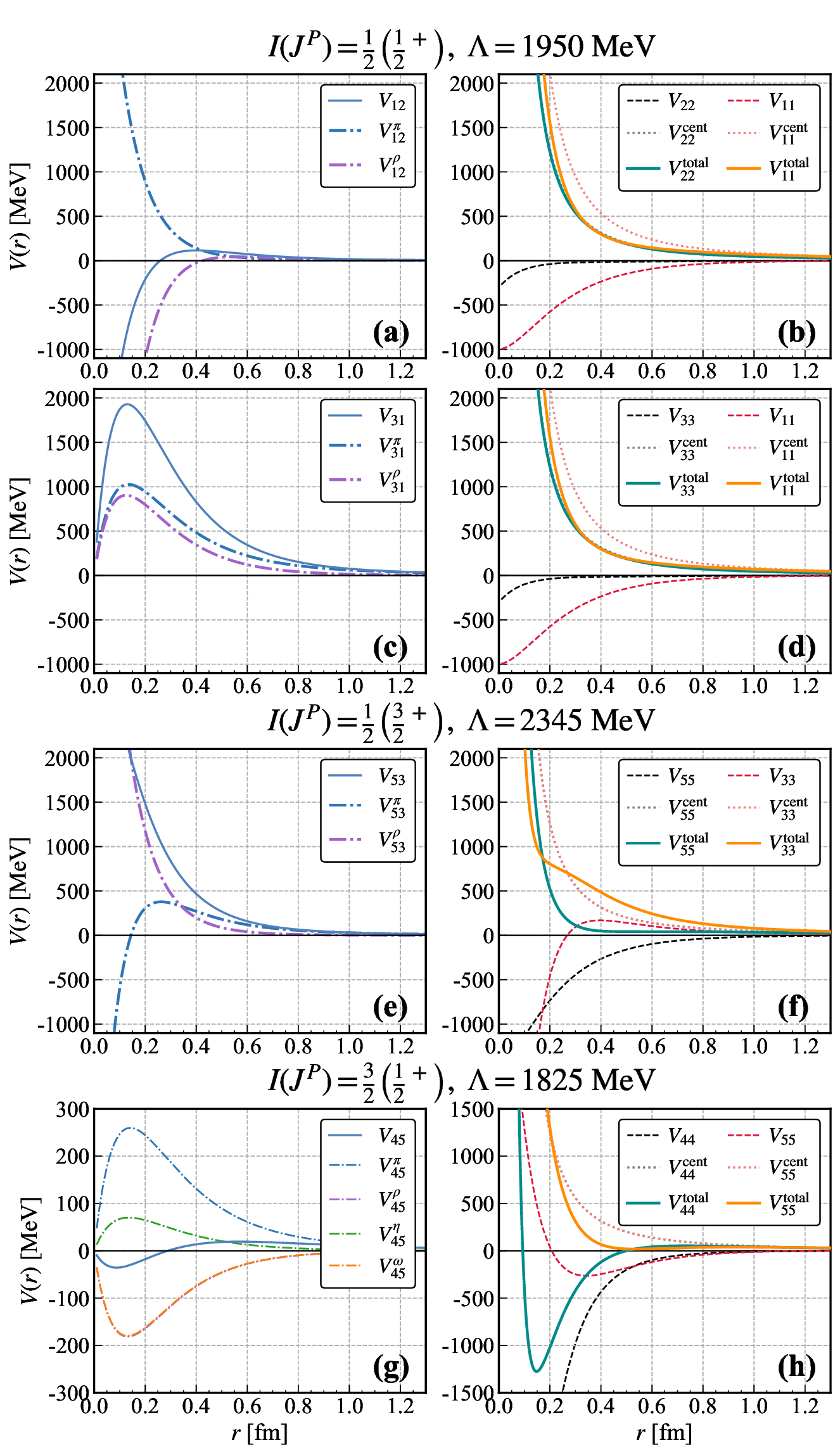}
\caption{Numerical results for the off-diagonal effective potentials (left column) and diagonal effective potentials (right column). The dashed, dotted, and solid curves in the right column represent the diagonal potentials $V_{ii}$, centrifugal barriers $V_{ii}^{\text{cent}}$, and total potentials $V_{ii}^{\text{total}}=V_{ii}+V_{ii}^{\text{cent}}$, respectively. The first two rows represent the $I(J^P)=1/2(1/2^+)$ system with $\Lambda=1950~\mathrm{MeV}$, the third row represents the $I(J^P)=1/2(3/2^+)$ system with $\Lambda=2345~\mathrm{MeV}$, and the fourth row represents the $I(J^P)=3/2(1/2^+)$ system with $\Lambda=1825~\mathrm{MeV}$. Channel indices $i\in\{1,2,3,4,5\}$ denote the $\Sigma_b K(^2P_J)$, $\Lambda_b K^*(^2P_J)$, $\Lambda_b K^*(^4P_J)$, $\Sigma_b K^*(^2P_J)$, and $\Sigma_b K^*(^4P_J)$ channels, respectively, with $J=1/2$ for $J^P=1/2^+$ and $J=3/2$ for $J^P=3/2^+$.}
\label{fig:Combined_Potentials_Paper_Final}
\end{figure*}

\subsubsection{$I(J^P)=1/2(1/2^+)$ $\Sigma_bK/\Lambda_bK^{*}/\Sigma_bK^{*}$}
First, we investigate the coupled $\Sigma_b K / \Lambda_b K^* / \Sigma_b K^*$ system with $I(J^P) = 1/2(1/2^+)$. Our numerical results show that, as the cutoff parameter $\Lambda$ increases, the pole undergoes a continuous evolution from a resonance to a bound state in the complex energy plane. At lower cutoff values, the state manifests as a narrow resonance located slightly above the $\Sigma_b K$ threshold. For instance, at $\Lambda = 1520$ MeV, the resonance is characterized by a
complex energy of $E-i\Gamma/2 = (6308.55-0.21i)$ MeV and a corresponding RMS radius $r_{\text{RMS}} = (1.14 + 0.58i)$ fm. As $\Lambda$ increases, the interaction strength is enhanced, and the pole moves below the threshold into the bound-state region. 

Furthermore, we analyze the channel proportions defined in Eq.~\eqref{eq:channel-cproduct}. Taking the state at $\Lambda = 1520$ MeV as a representative example, the $\Sigma_b K(^2P_{1/2})$, $\Lambda_b K^*(^2P_{1/2})$, $\Lambda_b K^*(^4P_{1/2})$, $\Sigma_b K^*(^2P_{1/2})$, and $\Sigma_b K^*(^4P_{1/2})$ channel proportions are $(51.31+4.14i)\%$, $(2.37-0.20i)\%$, $(28.95-2.45i)\%$, $(2.52-0.22i)\%$, and $(14.85-1.27i)\%$, respectively. This distribution remains stable as the cutoff varies. The state is predominantly composed of the $\Sigma_b K(^2P_{1/2})$, $\Lambda_b K^*(^4P_{1/2})$, and $\Sigma_b K^*(^4P_{1/2})$ components, indicating significant channel mixing and showing that the couplings among these channels play an important role in forming the state.

To further investigate the formation mechanism, we analyze various subsystems. In the single-channel calculations, no bound states or resonances are obtained, except for the $\Sigma_b K^*$ system at high cutoffs. Specifically, the $P$-wave centrifugal term $V^{\mathrm{cent}}_{ii}= l_i(l_i+1)/(2\mu_i r^2)$ acts as a repulsive barrier that counteracts binding. As illustrated in Figs.~\ref{fig:Combined_Potentials_Paper_Final}(b) and (d), for channels $i=1, 2, 3$, which represent $\Sigma_b K(^2P_{1/2})$, $\Lambda_b K^*(^2P_{1/2})$, and $\Lambda_b K^*(^4P_{1/2})$, respectively, the positive centrifugal barriers (dotted lines) exceed the magnitudes of the attractive diagonal effective potentials (dashed lines). Consequently, the overall diagonal interaction remains repulsive, precluding the emergence of states in the single-channel case. To generate a state, the system requires sufficient attraction to counteract the centrifugal barrier, which predominantly originates from channel couplings.

In the coupled subsystems, our calculations demonstrate that the $\Sigma_b K(^2P_{1/2})$, $\Lambda_b K^*(^4P_{1/2})$, and $\Sigma_b K^*(^4P_{1/2})$ channels are dominant. In the $\Lambda_b K^* / \Sigma_b K^*$ subsystem, a resonance is identified at $\Lambda = 1575$ MeV with a complex energy of $E-i\Gamma/2 = (6514.58- 0.22i)$ MeV and $r_{\text{RMS}} = (1.03 + 1.07i)$ fm. The $\Lambda_b K^*(^4P_{1/2})$ and $\Sigma_b K^*(^4P_{1/2})$ components dominate. In the $\Sigma_b K/\Lambda_b K^*$ subsystem, a resonance is found at $\Lambda=1950$ MeV with $E-i\Gamma/2=(6310.63-0.46i)$ MeV and $r_{\text{RMS}}=(1.06+0.71i)$ fm. In this state, the $\Sigma_b K(^2P_{1/2})$ and $\Lambda_b K^*(^4P_{1/2})$ channels dominate, with the sum of the real parts of their channel proportions exceeding 99\%, whereas the real part of the $\Lambda_b K^*(^2P_{1/2})$ channel proportion is less than 1\%.

To examine these coupling mechanisms, we focus on the $\Sigma_b K / \Lambda_b K^*$ subsystem and analyze its two-channel configurations. Specifically, when considering only the coupling between the two dominant channels, $\Sigma_b K(^2P_{1/2})$ and $\Lambda_b K^*(^4P_{1/2})$, a resonance is similarly obtained at $\Lambda = 1950$ MeV. The calculated self-energy is $\bar{\Sigma}_P^\theta = (-296.44 + 97.26i)$ MeV, with the $P$ and $Q$ spaces consisting of the $\Sigma_b K(^2P_{1/2})$ and $\Lambda_b K^*(^4P_{1/2})$ channels, respectively. Its negative real part indicates that this two-channel coupling induces an effective attraction. In contrast, no states are identified in the other two-channel configurations, namely $\Sigma_b K(^2P_{1/2})/\Lambda_b K^*(^2P_{1/2})$ and $\Lambda_b K^*(^2P_{1/2})/\Lambda_b K^*(^4P_{1/2})$. As illustrated in Figs.~\ref{fig:Combined_Potentials_Paper_Final}(a) and (c), this contrast arises because the off-diagonal potential $V_{13}$ exhibits significantly greater strength at long range than $V_{12}$, while $V_{23}$ is identically zero. Additionally, setting the $V_{13}$ coupling to zero precludes state formation in both the $\Sigma_b K(^2P_{1/2})/\Lambda_b K^*(^4P_{1/2})$ configuration and the $\Sigma_b K / \Lambda_b K^*$ subsystem. Consequently, these results confirm that the attraction in the $\Sigma_b K / \Lambda_b K^*$ subsystem predominantly originates from the $V_{13}$ coupling.

Furthermore, while resonant solutions can be formally obtained in the $\Sigma_b K^*$ subsystem or the coupled $\Sigma_b K / \Sigma_b K^*$ subsystem, the former requires a cutoff of $\Lambda = 3010$ MeV and the latter yields RMS radii $\operatorname{Re}(r_{\text{RMS}}) < 0.5$ fm. Such solutions fall outside the expected ranges of the cutoff and the RMS radius, thus they are not considered as molecular candidates.

\subsubsection{$I(J^P)=1/2(3/2^+)$ $\Sigma_bK/\Lambda_bK^{*}/\Sigma_bK^{*}$}
Similar to the $J^P = 1/2^+$ sector, a narrow resonance is identified in the $\Lambda_b K^* / \Sigma_b K^*$ system with $I(J^P) = 1/2(3/2^+)$ at $\Lambda = 2345$ MeV, with a complex energy of $E-i\Gamma/2= (6514.20-0.12i)$ MeV and an RMS radius $r_{\text{RMS}} = (1.03 + 1.09i)$ fm. This state originates entirely from coupled-channel effects, as single-channel calculations yield no solutions. The $\Lambda_b K^*(^4P_{3/2})$ and $\Sigma_b K^*(^4P_{3/2})$ channels dominate this state throughout the evolution, as listed in Table~\ref{tab:numerical_results}. As illustrated in Figs.~\ref{fig:Combined_Potentials_Paper_Final}(e) and (f), the sum of the diagonal potential ($V_{33}$ or $V_{55}$) and its corresponding centrifugal potential remains positive and therefore repulsive, whereas the off-diagonal potential $V_{35}$ contains a pion-exchange component that provides the long-range attraction and plays a key role in forming this resonance.

\subsubsection{$I(J^P)=3/2(1/2^+)$ $\Sigma_bK/\Sigma_bK^{*}$}
In the $I(J^P)=3/2(1/2^+)$ $\Sigma_b K^*$ subsystem, a resonance is obtained at $\Lambda = 1825$ MeV with $E-i\Gamma/2=(6707.14-0.30i)$ MeV and $r_{\text{RMS}} = (0.79 + 0.80i)$ fm. The state is dominated by the $\Sigma_b K^*(^2P_{1/2})$ component, with a channel proportion of $(97.98-0.73i)\%$. The formation of this state is attributed to the diagonal potential $V_{44}$ in Fig.~\ref{fig:Combined_Potentials_Paper_Final}(h), for which all meson-exchange terms are attractive. Conversely, the interaction potential $V_{55}$ for the $\Sigma_b K^*(^4P_{1/2})$ channel is repulsive, which strongly suppresses its contribution to the state. Notably, the potential $V_{44}^{\text{total}}$ shows a well followed by a barrier, indicating that this state exhibits features of a shape resonance.

\begin{figure*}[t!] 
\centering
\includegraphics[width=1.5\columnwidth]{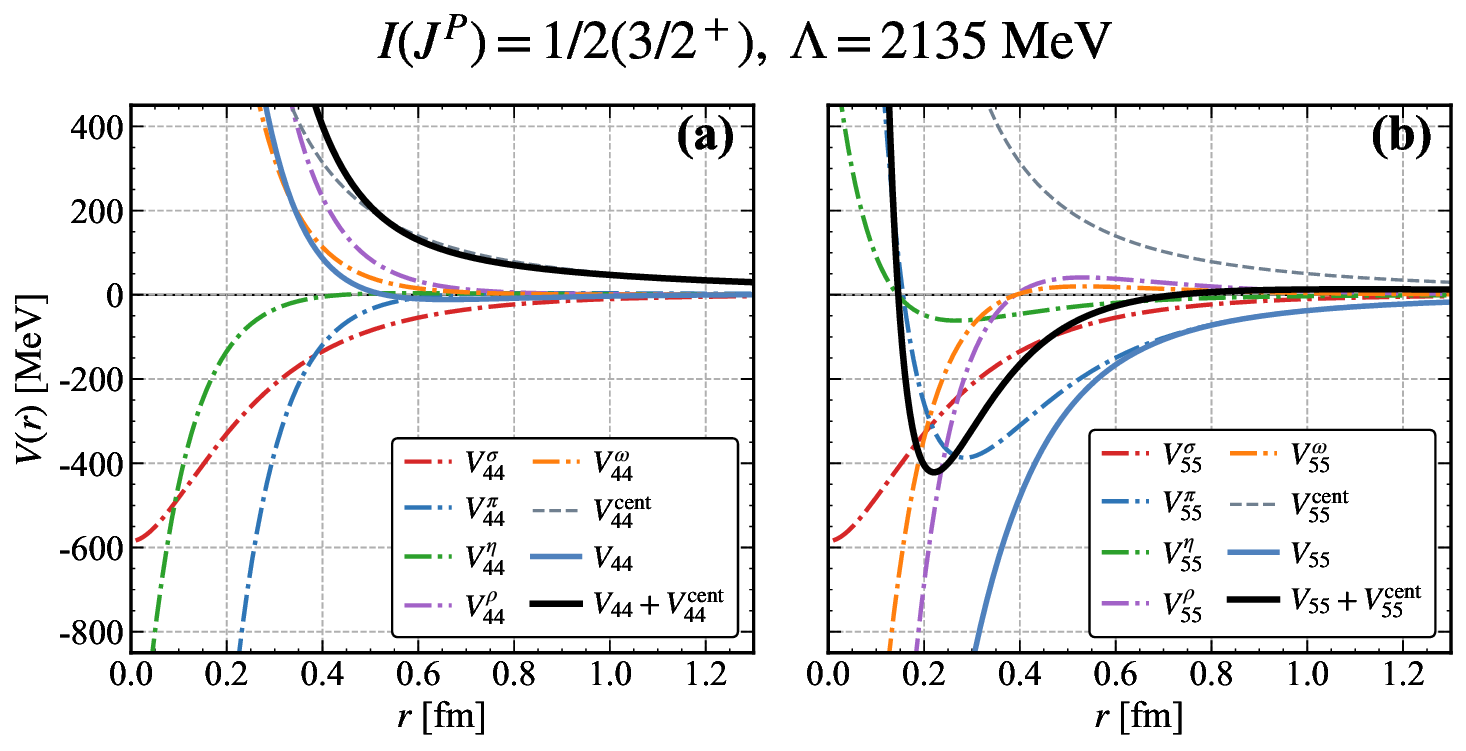}
 \caption{Potentials for the $\Sigma_b\bar{K}^{*}$ subsystem with $I(J^P)=1/2(3/2^+)$. Panels (a) and (b) display diagonal potentials $V_{ii}$, centrifugal barriers $V_{ii}^{\text{cent}}$, and their sums at $\Lambda = 2135$ MeV. Dot-dashed lines represent meson-exchange contributions. The indices $i = 4,5$ denote the $\Sigma_b \bar K^*(^2P_{3/2})$ and $\Sigma_b \bar K^*(^4P_{3/2})$ channels, respectively.}
 \label{fig:Potential_and_Trajectory_Combined_MeV}
\end{figure*}

\subsubsection{$I(J^P)=3/2(3/2^+)$ $\Sigma_bK/\Sigma_bK^{*}$}
The results for the $I(J^P)=3/2(3/2^+)$ system are analogous to the $I(J^P)=3/2(1/2^+)$ case. In the $\Sigma_b K^*$ subsystem, a resonance emerges at $\Lambda = 1825$ MeV with $E-i\Gamma/2 = (6707.97-0.42i)$ MeV and $r_{\text{RMS}} = (0.73 + 0.71i)$ fm. Similarly, the $\Sigma_b K^*(^2P_{3/2})$ component dominates with a channel proportion of $(99.46-0.31i)\%$, and the corresponding RMS radius remains small. This similarity to the $I(J^P)=3/2(1/2^+)$ system arises because the diagonal potential for the $\Sigma_b K^*(^2P_{3/2})$ channel is independent of the total angular momentum $J$. As in the $I(J^P)=3/2(1/2^+)$ system, the diagonal potential for the $\Sigma_b K^*(^4P_{3/2})$ channel is purely repulsive, rendering its contribution negligible.

\subsection{$\Lambda_b/\Sigma_b$ and $ \bar K^{(*)}$ interactions}
\subsubsection{$I(J^P)=1/2(1/2^+)$ $\Sigma_b\bar{K}/\Lambda_b\bar{K}^{*}/\Sigma_b\bar{K}^{*}$}
In the fully coupled $\Sigma_b \bar{K} / \Lambda_b \bar{K}^* / \Sigma_b \bar{K}^*$ system with $I(J^P) = 1/2(1/2^+)$, a resonance is found at $\Lambda = 2020$ MeV, characterized by a complex energy of $E-i\Gamma /2= (6310.42-0.39i)$ MeV and an RMS radius $r_{\text{RMS}} = (1.01 + 0.68i)$ fm. Similarly, for the $\Sigma_b \bar{K} / \Sigma_b \bar{K}^*$ subsystem, a narrow resonance emerges at $\Lambda = 2080$ MeV with $E-i\Gamma/2= (6308.53-0.15i)$ MeV and $r_{\text{RMS}} = (0.86 + 0.86i)$ fm. Single-channel calculations do not yield any states. The $\Sigma_b \bar{K}(^2P_{1/2})$ and $\Sigma_b \bar{K}^*(^4P_{1/2})$ components dominate in both coupled cases. In the fully coupled configuration, for instance, the real parts of their channel proportions sum to $94.26\%$, indicating that the formation mechanism is driven by channel coupling. These solutions exhibit a relatively strong sensitivity to the cutoff parameter, as shown in Table~\ref{tab:numerical_results}.

\subsubsection{$I(J^P)=1/2(3/2^+)$ $\Sigma_b\bar{K}/\Lambda_b\bar{K}^{*}/\Sigma_b\bar{K}^{*}$}
For the $I(J^P)=1/2(3/2^+)$ $\Sigma_b\bar{K}/\Lambda_b\bar{K}^{*}/\Sigma_b\bar{K}^{*}$ system, we first investigate the $\Sigma_b \bar{K}^*$ subsystem. When the $\Sigma_b \bar{K}^*(^2P_{3/2})$ and $\Sigma_b \bar{K}^*(^4P_{3/2})$ channels are coupled, a resonance above the $\Sigma_b \bar{K}^*$ threshold is obtained at $\Lambda = 2135$ MeV, with $E-i\Gamma /2= (6706.04-0.33i)$ MeV and $r_{\text{RMS}} = (1.24 + 1.41i)$ fm. The channel proportion of the $\Sigma_b \bar{K}^*(^4P_{3/2})$ component in this state is $(90.68+0.35i)\%$. As illustrated in Fig.~\ref{fig:Potential_and_Trajectory_Combined_MeV}(a), the sum of the diagonal potential for the $\Sigma_b \bar{K}^*(^2P_{3/2})$ channel and the $P$-wave centrifugal barrier remains entirely repulsive. Consequently, neither a bound state nor a resonance pole is found in the single-channel $\Sigma_b \bar{K}^*(^2P_{3/2})$ calculation.

In contrast, a resonance is found in the single-channel $\Sigma_b \bar{K}^*(^4P_{3/2})$ system. As shown in Fig.~\ref{fig:Potential_and_Trajectory_Combined_MeV}(b), the formation of this state is driven by a potential well in the total potential $V_{55}^{\text{total}}$, which is generated by the interplay between the $P$-wave centrifugal barrier and the $\pi$- and $\sigma$-exchange potentials. For $P$-wave systems, the centrifugal barrier suppresses the effects of the short-range $\rho/\omega$ interactions, allowing the long-range $\pi$-exchange and intermediate-range $\sigma$-exchange interactions to dominate the overall attraction~\cite{Machleidt:2017vls,Cheng:2026cgo}. Numerical calculations indicate that, in the absence of other meson-exchange contributions, the long-range attraction supplied by the $\pi$ exchange is sufficient to generate this resonance. This result aligns with the findings reported in Ref.~\cite{Wang:2024ukc}.

Furthermore, in the coupled $\Lambda_b \bar{K}^* / \Sigma_b \bar{K}^*$ system, a resonance emerges due to coupled-channel effects and can be regarded as a hadronic molecular candidate. As the cutoff varies from 2075 MeV to 2085 MeV, the complex energy evolves from $(6701.06 - 8.84i)$ MeV to $(6697.55 - 9.19i)$ MeV, and the RMS radius changes from $(1.05 - 0.33i)$ fm to $(1.01 - 0.24i)$ fm. This relatively broad resonance is dominated by the $\Sigma_b \bar{K}^*(^4P_{3/2})$ component, whose channel proportion is $(90.25+0.48i)\%$ at $\Lambda=2080$ MeV. The coupled-channel resonance evolves from the resonance in the $\Sigma_b \bar{K}^*$ subsystem as the $\Lambda_b\bar{K}^*$ channel is included. Compared with the resonance in the $\Sigma_b \bar{K}^*$ subsystem, the fully coupled state exhibits a lower energy and a broader width, suggesting that the coupled-channel interaction increases the effective attraction relative to the uncoupled $\Sigma_b \bar{K}^*$ subsystem. 

\subsubsection{$I(J^P)=3/2(1/2^+, 3/2^+)$ $\Sigma_b\bar{K}/\Sigma_b\bar{K}^{*}$}
For the $I(J^P)=3/2(1/2^+)$ system, a state is obtained only when the cutoff parameter $\Lambda$ approaches $4000$ MeV, which is excessively large. We therefore do not regard this solution as a credible hadronic molecular candidate. For the $I(J^P)=3/2(3/2^+)$ case, no bound state or resonance is found.

\section{Summary}\label{summary}

In this work, we systematically investigate the positive-parity $Y_bK^{(*)}$ and $Y_b\bar K^{(*)}$ systems within the framework of the OBE model. We explore all possible quantum number configurations spanning isospin $I=1/2,\, 3/2$ and spin-parity $J^P=1/2^+,\, 3/2^+,\, 5/2^+$. In contrast to the negative-parity sector investigated in our previous study~\cite{Song:2025yut}, where candidates were primarily identified for $\Lambda \in [800, 1100]$ MeV, the states found in the positive-parity sector are predominantly resonances and emerge at higher cutoff values ranging from 1500 to 2500 MeV. This behavior stems from the $P$-wave centrifugal barrier; the formation of these states requires stronger attraction to overcome the repulsive barrier, which consequently corresponds to higher cutoff values in the OBE model.

For the $Y_bK^{(*)}$ sector, our calculations identify several resonance poles. In the $I(J^P)=1/2(1/2^+)$ $\Sigma_b K/\Lambda_b K^*/\Sigma_b K^*$ system, we find a resonance candidate at $\Lambda = 1520$ MeV with $E-i\Gamma/2= (6308.55-0.21i)$ MeV and $r_{\text{RMS}} = (1.14+0.58i)$ fm. In the $I(J^P)=1/2(3/2^+)$ $\Lambda_b K^*/\Sigma_b K^*$ subsystem, a resonance candidate is identified at $\Lambda = 2345$ MeV with $E-i\Gamma/2 = (6514.20-0.12i)$ MeV and $r_{\text{RMS}} = (1.03+1.09i)$ fm. In the higher-isospin $I=3/2$ sector, two resonances are identified in the $\Sigma_b K^*$ subsystem: the $I(J^P)=3/2(1/2^+)$ state is obtained at $\Lambda = 1825$ MeV with $E-i\Gamma/2 = (6707.14-0.30i)$ MeV and $r_{\text{RMS}} = (0.79+0.80i)$ fm, and the $I(J^P)=3/2(3/2^+)$ state is found at $\Lambda = 1825$ MeV with $E-i\Gamma/2 = (6707.97-0.42i)$ MeV and $r_{\text{RMS}} = (0.73+0.71i)$ fm. The relatively small RMS radii of these $I=3/2$ resonances suggest that they are spatially more compact than the resonance candidates identified in the $I=1/2$ sector.

For the $Y_b\bar K^{(*)}$ sector, the fully coupled $I(J^P)=1/2(1/2^+)$ configuration yields a resonance at $\Lambda = 2020$ MeV with $E-i\Gamma/2 = (6310.42-0.39i)$ MeV and $r_{\text{RMS}} = (1.01+0.68i)$ fm, which exhibits strong cutoff sensitivity. In the $I(J^P)=1/2(3/2^+)$ configuration, a resonance is obtained in the $\Sigma_b\bar{K}^*$ subsystem at $\Lambda = 2135$ MeV with $E-i\Gamma/2 = (6706.04-0.33i)$ MeV and $r_{\text{RMS}} = (1.24+1.41i)$ fm. Upon introducing the $\Lambda_b\bar{K}^*$ coupling, a broad resonance candidate is obtained at $\Lambda = 2075$ MeV with $E-i\Gamma/2 = (6701.06 - 8.84i)$ MeV and $r_{\text{RMS}} = (1.05-0.33i)$ fm. In addition, the $I=3/2$ $\bar K^{(*)}$ sector yields no states compatible with a hadronic-molecular interpretation, and our investigation confirms that no bound states or resonances are found within the scanned cutoff range in any configuration with the highest spin $J=5/2$. We hope that the present predictions can provide useful guidance for the search for exotic bottom-baryon molecular states and contribute to further understanding of the nonperturbative dynamics of the strong interaction.

\clearpage

\appendix
\section{Numerical stability analysis}\label{App:further_discussion}

To verify the numerical reliability of the identified resonance poles, we perform a numerical stability analysis, taking the resonance in the $I(J^P)=1/2(3/2^+)$ $\Sigma_b \bar{K}^*$ system at $\Lambda = 2130$~MeV as a representative example. Figure~\ref{fig:stability_analysis} displays the dependence of the real part of the RMS radius, $\operatorname{Re}(r_{\text{RMS}})$, the energy $E$, and the decay width $\Gamma$ on the scaling angle $\theta$, the oscillator length $b_0$, and the number of basis functions $N$. As shown in the first column, $E$ and $\Gamma$ remain constant across the investigated $\theta$ range, while $\operatorname{Re}(r_{\text{RMS}})$ reaches a stable plateau for $\theta \ge 60^\circ$, consistent with the $\theta$-stability expected in the CSM. In the second column, the physical quantities stabilize over the range $b_0 \in [2.4, 4.0]$~GeV$^{-1}$. The third column shows that all quantities converge with respect to the number of basis functions and become stable for $N \ge 250$. Based on these stability tests, we adopt $b_0 = 3.0$~GeV$^{-1}$ and $N = 350$ within these stable regions for all systems, with $\theta = 60^\circ$ chosen in the plateau region for this state. These choices ensure that the calculated resonance properties are stable with respect to variations in the basis parameters and the scaling angle.

\onecolumngrid 

\begingroup
\setlength{\columnwidth}{\textwidth}
\begin{figure}[H] 
\centering
\includegraphics[width=0.88\textwidth]{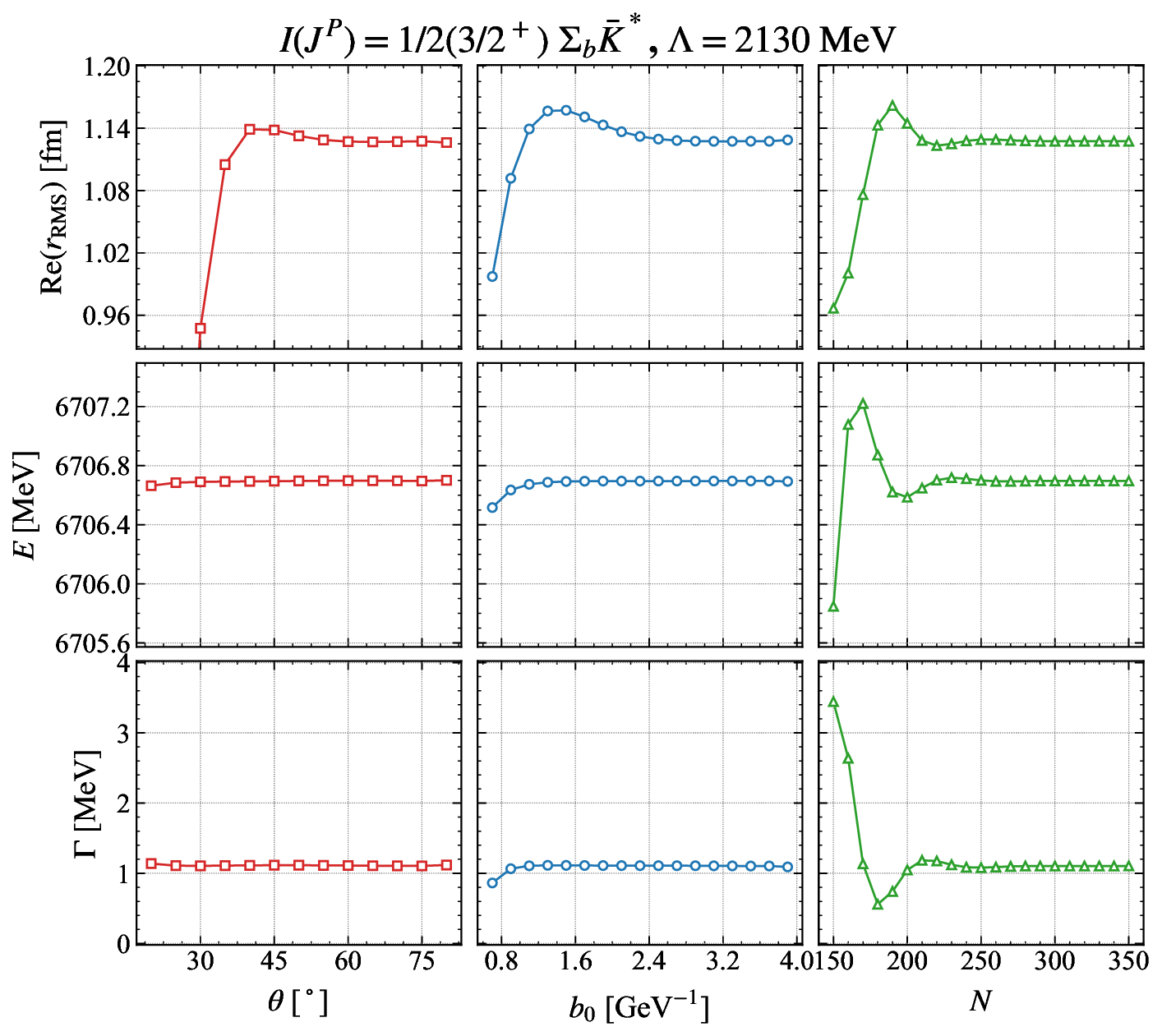}
\caption{The dependence of $\operatorname{Re}(r_{\text{RMS}})$, $E$, and $\Gamma$ on the scaling angle $\theta$, the oscillator length $b_0$, and the number of basis functions $N$ for the resonance in the $I(J^P)=1/2(3/2^+)$ $\Sigma_b \bar{K}^*$ system at $\Lambda = 2130$ MeV.}
\label{fig:stability_analysis}
\end{figure}
\endgroup

\twocolumngrid

\section{Explicit expressions of the scattering amplitudes}\label{App:Scattering_amplitude}
The total scattering amplitudes $\mathcal{M}$ for each process are listed in Table~\ref{amplitude_table}, which are the sum of the scattering amplitudes $\mathcal{M}_j$ correspond to each exchanged meson $j$ ($j=\sigma,\pi,\eta,\rho,\omega$). Here, $\mathcal{G}(I)$ represents the isospin factor, with $\mathcal{G}(1/2)=-2$ and $\mathcal{G}(3/2)=1$. $\mathcal{H}(\mathbf{q},\tilde{m}_j)\equiv 1/(\mathbf{q}^2+\tilde{m}_j^2)$, where $\mathbf{q}$ is the three-momentum transfer and $\tilde{m}_j$ denotes the effective mass of exchanged meson $j$. For process $h_1 h_2 \to h_3 h_4$ ($h_1h_2 \neq h_3h_4$), $m_{jk}$ denotes the effective mass $\tilde{m}_j$ of the exchanged meson $j$ in process $k$, with the energy transfer $q^0$ taken into account as in Eq.~\eqref{eq:m_eff}. For process $h_1 h_2 \to h_1 h_2$, $\tilde{m}_j$ is equivalent to the mass of exchanged meson $j$. $\chi_1$ and $\chi_3$ denote the two-component Pauli spinors of the initial and final baryons, while $\boldsymbol{\epsilon}_2$ and $\boldsymbol{\epsilon}_4$ denote the polarization vectors of the initial and final mesons, respectively. The amplitudes in the $Y_b\bar{K}^{(*)}$ sector can be obtained from the corresponding amplitudes in the $Y_bK^{(*)}$ sector using the $G$-parity rule~\cite{Klempt:2002ap}. Under this transformation, the amplitude associated with the exchange of meson $j$ transforms according to
\begin{align}
\mathcal{M}_{j}\to G_j\mathcal{M}_{j},
\qquad
G_j=\mathcal{C}_{j}(-1)^{I_j},
\end{align}
where $\mathcal{C}_{j}$ and $I_j$ are the $C$-parity and isospin, respectively. Because $G_\pi = G_\omega = -1$ and $G_\sigma = G_\rho = G_\eta = +1$, the amplitudes for the $Y_b\bar{K}^{(*)}$ sector are obtained by flipping the signs of the $\pi$- and $\omega$-exchange amplitudes in the $Y_bK^{(*)}$ sector, while the other amplitudes remain unchanged.

\onecolumngrid 

\renewcommand\arraystretch{2.4}
\setlength{\tabcolsep}{20pt}
\setlength{\LTpost}{0pt}
\setlength{\LTcapwidth}{\textwidth}

\begin{longtable}{c|c|l} 
\hline\hline
Process & $h_{1}h_{2}\to h_{3}h_{4}$ & $\mathcal{M}(h_{1}h_{2}\to h_{3}h_{4})$ \\
\hline
\endfirsthead

\endhead

\endfoot

\hline\hline
\caption{The total scattering amplitude $\mathcal{M}$ for each process}
\label{amplitude_table}
\endlastfoot

\multirow{2}{*}{1} & \multirow{2}{*}{$\displaystyle \Sigma_b K \to \Lambda_b K^*$}
& $\displaystyle - \frac{g_4 g}{f_{\pi}} \sqrt{m_{\Lambda_b} m_{\Sigma_b}} \chi_3^{\dagger}(\boldsymbol{\sigma}\cdot\mathbf{q})(\boldsymbol{\epsilon}_2\cdot \mathbf{q})\chi_1 \mathcal{H}(\mathbf{q}, m_{\pi1})$ \\
& & $\displaystyle - 2\sqrt{2}\lambda_I g_V g_{VVP} \sqrt{m_{\Lambda_b} m_{\Sigma_b}} m_{K^*} \chi_3^{\dagger}[(\boldsymbol{\sigma} \cdot \boldsymbol{\epsilon}_2)\mathbf{q}^2 - (\boldsymbol{\sigma} \cdot \mathbf{q})(\boldsymbol{\epsilon}_2 \cdot \mathbf{q})]\chi_1 \mathcal{H}(\mathbf{q}, m_{\rho1})$ \\
\hline

\multirow{3}{*}{2} & \multirow{3}{*}{$\displaystyle \Sigma_b K \to \Sigma_b K^*$}
& $\displaystyle \mathcal{G}(I) \frac{g_1 g}{\sqrt{2}f_{\pi}} m_{\Sigma_b} \chi_3^{\dagger}(\boldsymbol{\sigma}\cdot\mathbf{q})(\boldsymbol{\epsilon}_2 \cdot \mathbf{q})\chi_1 \mathcal{H}(\mathbf{q}, m_{\pi2}) + \frac{g_1 g}{3\sqrt{2} f_{\pi}} m_{\Sigma_b} \chi_3^{\dagger}(\boldsymbol{\sigma}\cdot\mathbf{q})(\boldsymbol{\epsilon}_2 \cdot \mathbf{q})\chi_1 \mathcal{H}(\mathbf{q}, m_{\eta2})$ \\
& & $\displaystyle + \mathcal{G}(I) \frac{2\lambda_S g_V g_{VVP}}{3} m_{\Sigma_b} m_{K^*} \chi_3^{\dagger}[(\boldsymbol{\sigma} \cdot \boldsymbol{\epsilon}_2) \mathbf{q}^2 - (\boldsymbol{\sigma} \cdot \mathbf{q}) (\boldsymbol{\epsilon}_2 \cdot \mathbf{q})]\chi_1 \mathcal{H}(\mathbf{q}, m_{\rho2})$ \\
& & $\displaystyle + \frac{2\lambda_S g_V g_{VVP}}{3} m_{\Sigma_b} m_{K^*} \chi_3^{\dagger}[(\boldsymbol{\sigma} \cdot \boldsymbol{\epsilon}_2) \mathbf{q}^2 - (\boldsymbol{\sigma} \cdot \mathbf{q}) (\boldsymbol{\epsilon}_2 \cdot \mathbf{q})]\chi_1 \mathcal{H}(\mathbf{q}, m_{\omega2})$ \\
\hline

\multirow{2}{*}{3} & \multirow{2}{*}{$\displaystyle \Lambda_b K^* \to \Sigma_b K^*$}
& $\displaystyle \frac{2g_4 g_{VVP}}{f_{\pi}} \sqrt{m_{\Lambda_b} m_{\Sigma_b}} m_{K^*} \chi_3^{\dagger}(\boldsymbol{\sigma}\cdot\mathbf{q}) [(\boldsymbol{\epsilon}_2 \times \boldsymbol{\epsilon}_4^*)\cdot\mathbf{q}]\chi_1 \mathcal{H}(\mathbf{q}, m_{\pi3})$ \\
& & $\displaystyle + \sqrt{2}\lambda_I g_V g \sqrt{m_{\Lambda_b} m_{\Sigma_b}} \chi_3^{\dagger} \left\{ [ \boldsymbol{\sigma} \cdot (\boldsymbol{\epsilon}_2 \times \boldsymbol{\epsilon}_4^*) ] \mathbf{q}^2 - (\boldsymbol{\sigma} \cdot \mathbf{q}) [(\boldsymbol{\epsilon}_2 \times \boldsymbol{\epsilon}_4^*)\cdot \mathbf{q}] \right\} \chi_1 \mathcal{H}(\mathbf{q}, m_{\rho3})$ \\
\hline

\multirow{2}{*}{4} & \multirow{2}{*}{$\displaystyle \Sigma_b K \to \Sigma_b K$}
& $\displaystyle -2 l_S g_\sigma m_{\Sigma_b} m_{K} (\chi_3^{\dagger} \chi_1) \mathcal{H}(\mathbf{q}, m_\sigma) + \mathcal{G}(I) \left(\frac{\beta_{S}g_Vg}{2} - \frac{\lambda_S g_Vg}{6m_{\Sigma_b}}\mathbf{q}^2\right) m_{\Sigma_b} m_K (\chi_3^{\dagger} \chi_1) \mathcal{H}(\mathbf{q}, m_\rho)$ \\
& & $\displaystyle + \left(\frac{\beta_{S}g_Vg}{2} - \frac{\lambda_S g_Vg}{6 m_{\Sigma_b}}\mathbf{q}^2\right) m_{\Sigma_b} m_K (\chi_3^{\dagger} \chi_1) \mathcal{H}(\mathbf{q}, m_\omega)$ \\
\hline

\multirow{1}{*}{5} & \multirow{1}{*}{$\displaystyle \Lambda_b K^* \to \Lambda_b K^*$}
& $\displaystyle 4 l_B g_{\sigma} m_{\Lambda_b} m_{K^*} (\chi_3^{\dagger} \chi_1) (\boldsymbol{\epsilon}_2 \cdot \boldsymbol{\epsilon}_4^{*}) \mathcal{H}(\mathbf{q}, m_\sigma)- \beta_B g_V g m_{\Lambda_b} m_{K^*} (\chi_3^{\dagger} \chi_1) (\boldsymbol{\epsilon}_2 \cdot \boldsymbol{\epsilon}_4^{*}) \mathcal{H}(\mathbf{q}, m_\omega)$ \\
\hline

\multirow{7}{*}{6} & \multirow{7}{*}{$\displaystyle \Sigma_b K^{*} \to \Sigma_b K^{*}$} 
& $\displaystyle -2 l_S g_{\sigma } m_{\Sigma_b} m_{K^*} (\chi_3^{\dagger} \chi_1) (\boldsymbol{\epsilon}_2 \cdot \boldsymbol{\epsilon}_4^{*}) \mathcal{H}(\mathbf{q}, m_\sigma)$ \\
& & $\displaystyle - \mathcal{G}(I) \frac{\sqrt{2}g_1 g_{VVP}}{f_{\pi}} m_{\Sigma_b} m_{K^*} \chi_3^{\dagger}(\boldsymbol{\sigma}\cdot\mathbf{q}) [(\boldsymbol{\epsilon}_2 \times \boldsymbol{\epsilon}_4^*)\cdot\mathbf{q}]\chi_1 \mathcal{H}(\mathbf{q}, m_\pi)$ \\
& & $\displaystyle - \frac{\sqrt{2}g_1 g_{VVP}}{3f_{\pi}} m_{\Sigma_b} m_{K^*} \chi_3^{\dagger}(\boldsymbol{\sigma}\cdot\mathbf{q}) [(\boldsymbol{\epsilon}_2 \times \boldsymbol{\epsilon}_4^*)\cdot\mathbf{q}]\chi_1 \mathcal{H}(\mathbf{q}, m_\eta)$ \\
& & $\displaystyle + \mathcal{G}(I) \left(\frac{\beta_{S}g_Vg}{2} - \frac{\lambda_S g_Vg}{6m_{\Sigma_b}}\mathbf{q}^2\right) m_{\Sigma_b} m_{K^*} (\chi_3^{\dagger} \chi_1) (\boldsymbol{\epsilon}_2\cdot\boldsymbol{\epsilon}_4^{*}) \mathcal{H}(\mathbf{q}, m_\rho)$ \\
& & $\displaystyle - \mathcal{G}(I) \frac{\lambda_S g_V g}{3} m_{\Sigma_b} \chi_3^{\dagger}\left\{ [ \boldsymbol{\sigma} \cdot (\boldsymbol{\epsilon}_2 \times \boldsymbol{\epsilon}_4^*) ] \mathbf{q}^2 - (\boldsymbol{\sigma} \cdot \mathbf{q}) [(\boldsymbol{\epsilon}_2 \times \boldsymbol{\epsilon}_4^*)\cdot\mathbf{q}] \right\} \chi_1 \mathcal{H}(\mathbf{q}, m_\rho)$ \\
& & $\displaystyle + \left(\frac{\beta_{S}g_Vg}{2} - \frac{\lambda_S g_Vg}{6 m_{\Sigma_b}}\mathbf{q}^2\right) m_{\Sigma_b} m_{K^*} (\chi_3^{\dagger} \chi_1) (\boldsymbol{\epsilon}_2\cdot\boldsymbol{\epsilon}_4^{*}) \mathcal{H}(\mathbf{q}, m_\omega)$ \\
& & $\displaystyle - \frac{\lambda_S g_V g}{3} m_{\Sigma_b} \chi_3^{\dagger}\left\{ [ \boldsymbol{\sigma} \cdot (\boldsymbol{\epsilon}_2 \times \boldsymbol{\epsilon}_4^*) ] \mathbf{q}^2 - (\boldsymbol{\sigma} \cdot \mathbf{q}) [(\boldsymbol{\epsilon}_2 \times \boldsymbol{\epsilon}_4^*)\cdot\mathbf{q}] \right\} \chi_1 \mathcal{H}(\mathbf{q}, m_\omega)$ 

\end{longtable}

\twocolumngrid


\begin{thebibliography}{99}
\bibitem{Belle:2003nnu}
S.~K.~Choi \textit{et al.} [Belle],
Phys. Rev. Lett. \textbf{91}, 262001 (2003)

\bibitem{Lee:2009hy}
I.~W.~Lee, A.~Faessler, T.~Gutsche and V.~E.~Lyubovitskij,
Phys. Rev. D \textbf{80}, 094005 (2009)

\bibitem{Liu:2019zoy}
Y.~R.~Liu, H.~X.~Chen, W.~Chen, X.~Liu and S.~L.~Zhu,
Prog. Part. Nucl. Phys. \textbf{107}, 237-320 (2019)

\bibitem{Esposito:2025hlp}
A.~Esposito, A.~Glioti, D.~Germani and A.~D.~Polosa,
Riv. Nuovo Cim. \textbf{48}, no.2, 95-155 (2025)

\bibitem{Dong:2017gaw}
Y.~Dong, A.~Faessler and V.~E.~Lyubovitskij,
Prog. Part. Nucl. Phys. \textbf{94}, 282-310 (2017)

\bibitem{Guo:2017jvc}
F.~K.~Guo, C.~Hanhart, U.~G.~Mei{\ss}ner, Q.~Wang, Q.~Zhao and B.~S.~Zou,
Rev. Mod. Phys. \textbf{90}, no.1, 015004 (2018)
[erratum: Rev. Mod. Phys. \textbf{94}, no.2, 029901 (2022)]

\bibitem{Zou:2013af}
B.~S.~Zou,
Nucl. Phys. A \textbf{914}, 454-460 (2013)

\bibitem{Zou:2021sha}
B.~S.~Zou,
Sci. Bull. \textbf{66}, 1258 (2021)

\bibitem{Karliner:2017qhf}
M.~Karliner, J.~L.~Rosner and T.~Skwarnicki,
Ann. Rev. Nucl. Part. Sci. \textbf{68}, 17-44 (2018)

\bibitem{Chen:2022asf}
H.~X.~Chen, W.~Chen, X.~Liu, Y.~R.~Liu and S.~L.~Zhu,
Rept. Prog. Phys. \textbf{86}, no.2, 026201 (2023)

\bibitem{LHCb:2020bls}
R.~Aaij \textit{et al.} [LHCb],
Phys. Rev. Lett. \textbf{125}, 242001 (2020)

\bibitem{LHCb:2020pxc}
R.~Aaij \textit{et al.} [LHCb],
Phys. Rev. D \textbf{102}, 112003 (2020)

\bibitem{Chen:2020aos}
H.~X.~Chen, W.~Chen, R.~R.~Dong and N.~Su,
Chin. Phys. Lett. \textbf{37}, no.10, 101201 (2020)

\bibitem{He:2020btl}
J.~He and D.~Y.~Chen,
Chin. Phys. C \textbf{45}, no.6, 063102 (2021)

\bibitem{Burns:2020epm}
T.~J.~Burns and E.~S.~Swanson,
Phys. Lett. B \textbf{813}, 136057 (2021)

\bibitem{Agaev:2020nrc}
S.~S.~Agaev, K.~Azizi and H.~Sundu,
J. Phys. G \textbf{48}, no.8, 085012 (2021)

\bibitem{Xiao:2020ltm}
C.~J.~Xiao, D.~Y.~Chen, Y.~B.~Dong and G.~W.~Meng,
Phys. Rev. D \textbf{103}, no.3, 034004 (2021)

\bibitem{Ke:2022ocs}
H.~W.~Ke, Y.~F.~Shi, X.~H.~Liu and X.~Q.~Li,
Phys. Rev. D \textbf{106}, no.11, 114032 (2022)

\bibitem{Wang:2024ukc}
J.~Z.~Wang, Z.~Y.~Lin, B.~Wang, L.~Meng and S.~L.~Zhu,
Phys. Rev. D \textbf{110}, no.11, 114003 (2024)

\bibitem{Lu:2026klm}
Q.~Lu, Z.~Zhu, C.~Cheng and Y.~Huang,

\bibitem{Chen:2023qlx}
R.~Chen and Q.~Huang,
Phys. Rev. D \textbf{108}, no.5, 054011 (2023)

\bibitem{Sheng:2024hkf}
L.~C.~Sheng, J.~Y.~Huo, R.~Chen, F.~L.~Wang and X.~Liu,
Phys. Rev. D \textbf{110}, no.5, 054044 (2024)

\bibitem{Yan:2026ryi}
Y.~Yan, Q.~Huang, Y.~Wu, H.~Huang and J.~Ping,

\bibitem{Song:2025yut}
Q.~F.~Song, Q.~F.~L{\"u} and X.~Xiong,
Eur. Phys. J. C \textbf{85}, no.9, 1026 (2025)

\bibitem{LHCb:2018vuc}
R.~Aaij \textit{et al.} [LHCb],
Phys. Rev. Lett. \textbf{121}, no.7, 072002 (2018)

\bibitem{Wan:2026xzg}
Y.~X.~Wan, R.~Chen, F.~L.~Wang and Q.~Huang,
[arXiv:2607.14913 [hep-ph]].

\bibitem{Song:2025ijd}
Q.~F.~Song, W.~Liang, Q.~F.~L{\"u} and X.~Xiong,
Phys. Rev. D \textbf{113}, no.5, 054001 (2026)

\bibitem{Su:2025toa}
J.~C.~Su, Q.~F.~Song, Q.~F.~L{\"u} and J.~Zhu,
Eur. Phys. J. C \textbf{85}, no.10, 1181 (2025)

\bibitem{Song:2024ngu}
Q.~F.~Song, Q.~F.~L{\"u}, D.~Y.~Chen and Y.~B.~Dong,
Phys. Rev. D \textbf{110}, no.7, 074038 (2024)

\bibitem{Liu:2018bkx}
M.~Z.~Liu, T.~W.~Wu, J.~J.~Xie, M.~Pavon Valderrama and L.~S.~Geng,
Phys. Rev. D \textbf{98}, no.1, 014014 (2018)

\bibitem{Machleidt:2017vls}
R.~Machleidt,
Int. J. Mod. Phys. E \textbf{26}, no.11, 1730005 (2017)

\bibitem{Liu:2011xc}
Y.~R.~Liu and M.~Oka,
Phys. Rev. D \textbf{85}, 014015 (2012)

\bibitem{Lin:1999ad}
Z.~w.~Lin and C.~M.~Ko,
Phys. Rev. C \textbf{62}, 034903 (2000)

\bibitem{Chen:2017xat}
R.~Chen, A.~Hosaka and X.~Liu,
Phys. Rev. D \textbf{97}, no.3, 036016 (2018)

\bibitem{Kaymakcalan:1983qq}
O.~Kaymakcalan, S.~Rajeev and J.~Schechter,
Phys. Rev. D \textbf{30}, 594 (1984)

\bibitem{ParticleDataGroup:2026mpi}
F.~Takahashi \textit{et al.} [Particle Data Group],
Int. J. Mod. Phys. A \textbf{41}, no.22, 2630011 (2026)

\bibitem{Breit:1929zz}
G.~Breit,
Phys. Rev. \textbf{34}, 553-573 (1929)

\bibitem{Breit:1930zza}
G.~Breit,
Phys. Rev. \textbf{36}, 383-397 (1930)

\bibitem{Machleidt:1987hj}
R.~Machleidt, K.~Holinde and C.~Elster,
Phys. Rept. \textbf{149}, 1-89 (1987)

\bibitem{Li:2012bt}
N.~Li and S.~L.~Zhu,
Phys. Rev. D \textbf{86}, 014020 (2012)

\bibitem{Aguilar:1971ve}
J.~Aguilar and J.~M.~Combes,
Commun. Math. Phys. \textbf{22}, 269-279 (1971)

\bibitem{Balslev:1971vb}
E.~Balslev and J.~M.~Combes,
Commun. Math. Phys. \textbf{22}, 280-294 (1971)

\bibitem{Simon:1972qft}
B.~Simon,
Commun. Math. Phys. \textbf{27}, no.1, 1-9 (1972)

\bibitem{Reinhardt:1982com}
W.~P.~Reinhardt,
Ann. Rev. Phys. Chem. \textbf{33}, no.1, 223-255 (1982)

\bibitem{Ho:1983lwa}
Y.~K.~Ho,
Phys. Rept. \textbf{99}, no.1, 1-68 (1983)

\bibitem{Moiseyev:1998gjp}
N.~Moiseyev,
Phys. Rept. \textbf{302}, no.5-6, 212-293 (1998)

\bibitem{Myo:2014ypa}
T.~Myo, Y.~Kikuchi, H.~Masui and K.~Kat{\={o}},
Prog. Part. Nucl. Phys. \textbf{79}, 1-56 (2014)

\bibitem{Yu:2021lmb}
Z.~Yu, M.~Song, J.~Y.~Guo, Y.~Zhang and G.~Li,
Phys. Rev. C \textbf{104}, no.3, 035201 (2021)

\bibitem{Yamani:1975jmm}
H.~A.~Yamani and L.~Fishman,
J. Math. Phys. \textbf{16}, no.2, 410-420 (1975)

\bibitem{Wang:2021hho}
C.~H.~Wang, L.~Tang, T.~Y.~Li, G.~P.~Zheng, J.~F.~Hu and C.~Q.~Pang,
Nucl. Phys. Rev. \textbf{39}, no.2, 160-171 (2022)

\bibitem{Sharaf:2019jgw}
M.~Sharaf, R.~McCarty, R.~A.~M.~Basili and J.~P.~Vary,

\bibitem{Caprio:2012as}
M.~A.~Caprio, P.~Maris and J.~P.~Vary,
J. Phys. Conf. Ser. \textbf{403}, 012014 (2012)

\bibitem{Homma:1997wtc}
M.~Homma, T.~Myo and K.~Kato,
Prog. Theor. Phys. \textbf{97}, no.4, 561-567 (1997)

\bibitem{Lin:2023ihj}
Z.~Y.~Lin, J.~B.~Cheng, B.~L.~Huang and S.~L.~Zhu,
Phys. Rev. D \textbf{108}, no.11, 114014 (2023)

\bibitem{Moiseyev:2011}
N.~Moiseyev, 
\textit{Non-Hermitian Quantum Mechanics} (Cambridge University Press, 2011).



\bibitem{Berggren:1970wto}
T.~Berggren,
Phys. Lett. B \textbf{33}, 547-549 (1970)

\bibitem{Gyarmati:1972yac}
B.~Gyarmati, F.~Krisztinkovics and T.~Vertse,
Phys. Lett. B \textbf{41}, 110-112 (1972)

\bibitem{Myo:2023heu}
T.~Myo and K.~Kat{\={o}},
Phys. Rev. C \textbf{107}, no.1, 014301 (2023)

\bibitem{Feshbach:1958nx}
H.~Feshbach,
Annals Phys. \textbf{5}, 357-390 (1958)

\bibitem{Feshbach:1962ut}
H.~Feshbach,
Annals Phys. \textbf{19}, 287-313 (1962)

\bibitem{Saito:1994iz}
T.~Y.~Saito and I.~R.~Afnan,
Few Body Syst. \textbf{18}, 101-132 (1995)

\bibitem{Yamaguchi:2017zmn}
Y.~Yamaguchi, A.~Giachino, A.~Hosaka, E.~Santopinto, S.~Takeuchi and M.~Takizawa,
Phys. Rev. D \textbf{96}, no.11, 114031 (2017)

\bibitem{Tornqvist:1993ng}
N.~A.~Tornqvist,
Z. Phys. C \textbf{61}, 525-537 (1994)

\bibitem{Tornqvist:1993vu}
N.~A.~Tornqvist,
Nuovo Cim. A \textbf{107}, 2471-2476 (1994)

\bibitem{Yang:2011wz}
Z.~C.~Yang, Z.~F.~Sun, J.~He, X.~Liu and S.~L.~Zhu,
Chin. Phys. C \textbf{36}, 6-13 (2012)

\bibitem{Cheng:2026cgo}
J.~B.~Cheng, Z.~Y.~Lin, J.~Z.~Wang and S.~L.~Zhu,
Phys. Rev. D \textbf{113}, no.9, 096001 (2026)

\bibitem{Klempt:2002ap}
E.~Klempt, F.~Bradamante, A.~Martin and J.~M.~Richard,
Phys. Rept. \textbf{368}, 119-316 (2002)
\end{thebibliography}
\end{document}